\documentclass[%
 aip,
 amsmath,amssymb,
preprint,%
]{revtex4-1}

\usepackage{graphicx}
\usepackage{dcolumn}
\usepackage{bm}

\usepackage[utf8]{inputenc}
\usepackage[T1]{fontenc}
\usepackage{etoolbox}

\usepackage{savesym}
\savesymbol{ao}
\savesymbol{vr}
\usepackage[notes]{ctmmath-v3} 
\usepackage{threeparttable}

\newcommand{\dsgiw}{\dot{\gkg}_w}
\newcommand{\hmuiw}{\hat{\mu}_w}

\newcommand{\tauiw}{\ten\tau_w}

\newcommand{\Cisw}{C_{iw}}
\newcommand{\Ciswo}{\overline{C}_{iw}}

\newcommand{\CiAw}{C_{Aw}}

\newcommand{\uiw}{\vec u_{iw}}

\begin{document}

\title{Semi-analytical solutions of passive scalar transport in generalized Newtonian fluid flow}
\author{Christopher A. Bowers}
\email{cabower2@ncsu.edu}
\affiliation{The University of North Carolina at Chapel Hill}
\affiliation{North Carolina State University}
\author{Cass T. Miller}
\affiliation{The University of North Carolina at Chapel Hill}

\date{\today}

\begin{abstract}

Transport during flow of generalized Newtonian fluids (GNFs) appears often in systems that can be treated in a simplified form as either cylindrical tubes or slit openings between parallel plates. Based on the pioneering work of Taylor, analytical solutions for transport in these simplified systems were derived generally. This includes analytical solutions for advection dominated transport, as well as a computation of the enhanced molecular diffusion coefficient in low Peclet number systems. The newly derived general solutions for species transport were applied to Cross and Carreau model fluids using a semi-analytical solution for velocity of these fluids. The semi-analytical solutions derived herein were compared to microscale simulations and showed agreement to within the numerical error of those simulations. The semi-analytical transport solutions derived here were developed without assuming any specific fluid rheology, thus these solutions can be applied to other non-Newtonian fluids, such as viscoelastic or viscoplastic fluids, as a straightforward extension of this work. 

\end{abstract}

\maketitle

\section{Introduction}
\label{sec:introduction}

The solution of flow and transport in capillary tubes and the slits formed between two parallel plates are classic problems in fluid mechanics that are regularly studied due to their relative simplicity, as well as their applicability to several real world systems. Famously, Taylor \cite{Taylor_53} derived an analytical solution for the average concentration at different locations in a tube during advection dominated species transport of a passive scalar. Taylor also derived a relationship for species dispersion in a capillary tube when advection is dominant in the longitudinal direction but diffusion is dominant in the transverse direction, a phenomenon known as enhanced molecular diffusion \cite{Fan_Hwang_65,Agrawal_Jayaraman_etal_93,Rosencrans_97,MacDonald_Price_etal_19,Taylor_Harris_19,Bairwa_Khosa_etal_23}. Aris \cite{Aris_56} then derived a more general solution that could apply to various viscous flows in different conduit cross-sections using the method of moments. Since these works, there has been special interest in species transport and enhanced molecular diffusion during non-Newtonian fluid flow in different conduits \cite{Zhang_Frigaard_06,Garg_Prasad_24}, including in capillary tubes and slits \cite{Boschan_Ippolito_etal_08,Zhang_Prodanovic_etal_19,Rana_Liao_19,Shende_Niasar_etal_21,Brown_Dejam_23}. However, past derivations have relied on closed-form solutions for the velocity profile within the geometry of interest, which is not accessible for many non-Newtonian fluids, including Cross and Carreau model fluids \cite{Sochi_15,Hauswirth_Bowers_etal_20}. A general solution is needed for species transport in reference geometries that does not rely on the existence of a closed-form solution for velocity. 

Non-Newtonian fluids are fluids that do not conform to Newton's law of viscosity, which describes a linear relationship between shear stress and shear rate \cite{Bird_Stewart_etal_02}. There are a wide variety of non-Newtonian fluids.  A subset of non-Newtonian fluids of interest are generalized Newtonian fluids (GNFs), which appear routinely in industrial and biological settings \cite{Sorbie_Clifford_etal_89,Pearson_Tardy_02,Sochi_10,Zhang_Prodanovic_etal_19,Hauswirth_Bowers_etal_20,Bowers_Miller_21,Bowers_Miller_23,Bowers_Miller_25}.GNFs follow a generalized form of Newton's law of viscosity, given by
\begin{equation}
    \tauiw = 2\hmuiw\lrp{\dsgiw}\diw\;,
    \label{eq:tauiw}
\end{equation}
where $\tauiw$ is the viscous stress tensor; $\hmuiw$ is the dynamic viscosity, which depends on shear rate $\dsgiw$; $\diw$ is the symmetric rate of strain tensor; $w$ denotes the liquid phase; and subscripts indicate that these are fluid-continuum scale quantities where no averaging has occurred to a larger scale\cite{Skelland_67,Bird_Stewart_etal_02,Gray_Miller_14,Bowers_Miller_21}. The rate of strain tensor is defined as 
\begin{equation}
    \diw = \frac 12\lrb{\del\viw + \lrp{\del\viw}\T}\;,
    \label{eq:diw}
\end{equation}
where $\viw$ is the flluid velocity \cite{Bird_Stewart_etal_02,Gray_Miller_14}. The shear rate is defined by \cite{Bird_Stewart_etal_02,Hauswirth_Bowers_etal_20}
\begin{equation}
    \dsgiw = \sqrt{2\diw\dd\diw}\;.
    \label{eq:dsgiw}
\end{equation}

Two classes of GNFs of particular interest are Cross and Carreau model fluids \cite{Cross_65,Tosco_Marchisio_etal_13,Zhang_Prodanovic_etal_19,Hauswirth_Najm_etal_19,Basset_Najm_etal_19,Perrin_Tardy_etal_06,Sochi_10,Sorbie_Clifford_etal_89,Tosco_Marchisio_etal_13,Zami-Pierre_Loubens_etal_18,Castro_Agnaou_19,Castro_Goyeau_21}, which both exhibit infinite- and zero-shear limit Newtonian viscosities, $\hsmu_\infty$ and $\hsmu_0$ respectively, as well as a non-Newtonian behavior index $n$ and a fourth parameter $m$. The Cross model is often used to describe hydraulic fracturing fluids \cite{Barbati_Desroches_etal_16,Barati_Liang_14,Norman_Jasinski_etal_96,Buchley_Lord_73}, while the Carreau model has been used to describe both polymers for subsurface operations \cite{Gerrard_Perutz_etal_52,Ancey_Meunier_04,Ancey_07,Barr_Pappalardo_etal_04,Hulme_74,Wu_02,Sonder_Zimanowski_etal_06,Lev_Spiegelman_etal_12} and the rheology of biological fluids \cite{Peyrounette_Davit_etal_18,Bessonov_Simakov_etal_16,Sriram_Intaglietta_etal_14,Chakraborty_05,Rabby_Razzak_etal_13,Zami-Pierre_Loubens_etal_18}. The Cross model can be written as \cite{Cross_65,Sochi_10}
\begin{equation}
    \hmuiw\lrp{\dsgiw} = \hsmu_\infty + \frac{\hsmu_0-\hsmu_\infty}{1+\lrp{m\dsgiw}^n}\;,
    \label{eq:Cross}
\end{equation}
while the Carreau model can be formulated as \cite{Sochi_15,Bowers_Miller_23}
\begin{equation}
    \hmuiw\lrp{\dsgiw} = \hsmu_\infty + \frac{\hsmu_0-\hsmu_\infty}{\lrb{1+\lrp{m\dsgiw}^2}^{\frac{1-n}{2}}}\;.
    \label{eq:Carreau}
\end{equation}

The Cross and Carreau models have proven challenging to treat analytically for most reference geometries, including capillary tubes and slits. In such geometries, the shear stress is computed directly from a force balance, and then the relationship between shear stress and shear rate is used to compute the velocity via a method that has become known as the Weissenberg-Rabinowitsch-Mooney and Schofield (WRMS) method \cite{Skelland_67,Sochi_15,Kim_18,Wrobel_20}. Because there is no closed-form relationship between shear rate and shear stress, as can be seen by substituting \Eqn{Cross} or \Eqn{Carreau} into \Eqn{tauiw}, the WRMS method also does not result in a closed-form solution. Open-form analytical solutions for Cross and Carreau model fluids have been proposed where the unknown shear rates are computed numerically, including for flow rate \cite{Sochi_15} and velocity \cite{Kim_18,Wrobel_20,Wang_22} in capillary tubes. Special attention should be paid to the work of Kim \cite{Kim_18}, which is not well known in the literature. While solutions to velocity are necessary to generate species transport models, these open-form solutions cannot be readily adapted to current theory. 

Species transport during GNF flow in capillary tubes is of interest to the medical field as a simplified analogue of blood vessels \cite{Taylor_53,Sirs_91,Rana_Liao_19,Kutev_Tobakova_etal_21}, as well as to industrial applications where polymers are pumped through piping \cite{Pinho_Whitelaw_90,Guzel_Frigaard_etal_09,Ryltseva_Borzenko_etal_20,Shende_Niasar_etal_21}. The parallel plate geometry is of special interest to industries that operate in the subsurface because it is often used as a simplified analogue of fractures in media \cite{Boschan_Ippolito_etal_08,Sochi_15,Zhang_Prodanovic_etal_19}. There has been numerical and experimental investigation of species transport during GNF flows in different conduits \cite{Zhang_Frigaard_06,Sochi_15}, and theoretical investigation into flow of fluids exhibiting other rheologies in capillary tubes \cite{Pantokratoras_16,Kim_18,Wrobel_20,Wang_22}; however, a theoretical treatment of species transport of a passive scalar in Cross and Carreau model fluid flow in capillary tubes and slits has not appeared in the literature. 

The goal of this work is to develop semi-analytical solutions for flow and passive scalar transport of Cross and Carreau model fluids in capillary tubes and slits formed between parallel plates. The specific objectives are to: (1) derive a semi-analytical solution for velocity; (2) derive an analytical solution for species concentration when velocities are known during advection-dominated flow; (3) derive an approximate analytical solution for species concentration during enhanced molecular diffusion; (4) validate the semi-analytical solutions derived herein by comparison to numerical simulations; and (5) discuss extensions of the results to other classes of non-Newtonian fluids.

\section{Analytical Solutions for Velocity}
\label{sec:analyticV}

Any analytical solution for species transport will require knowledge of the velocity field within a system. Thus, analytical solutions for velocity will be derived first, followed by species transport solutions. While there are existing solutions to velocity in a capillary tube for Cross and Carreau model fluids \cite{Kim_18,Wang_22}, none appear in the literature for slits between parallel plates, and existing solutions often use various complex mathematical formulations for their derivation. \cite{Wrobel_20,Shende_Niasar_etal_21,Pricci_Tullio_etal_22} A simplified derivation, based on the WRMS method,\cite{Skelland_67,Sochi_15,Kim_18} will be presented here that applies for both capillary tubes and slits. 

\subsection{General Solution}

\begin{figure}[t!]
   \centering
    \includegraphics[width=\linewidth]{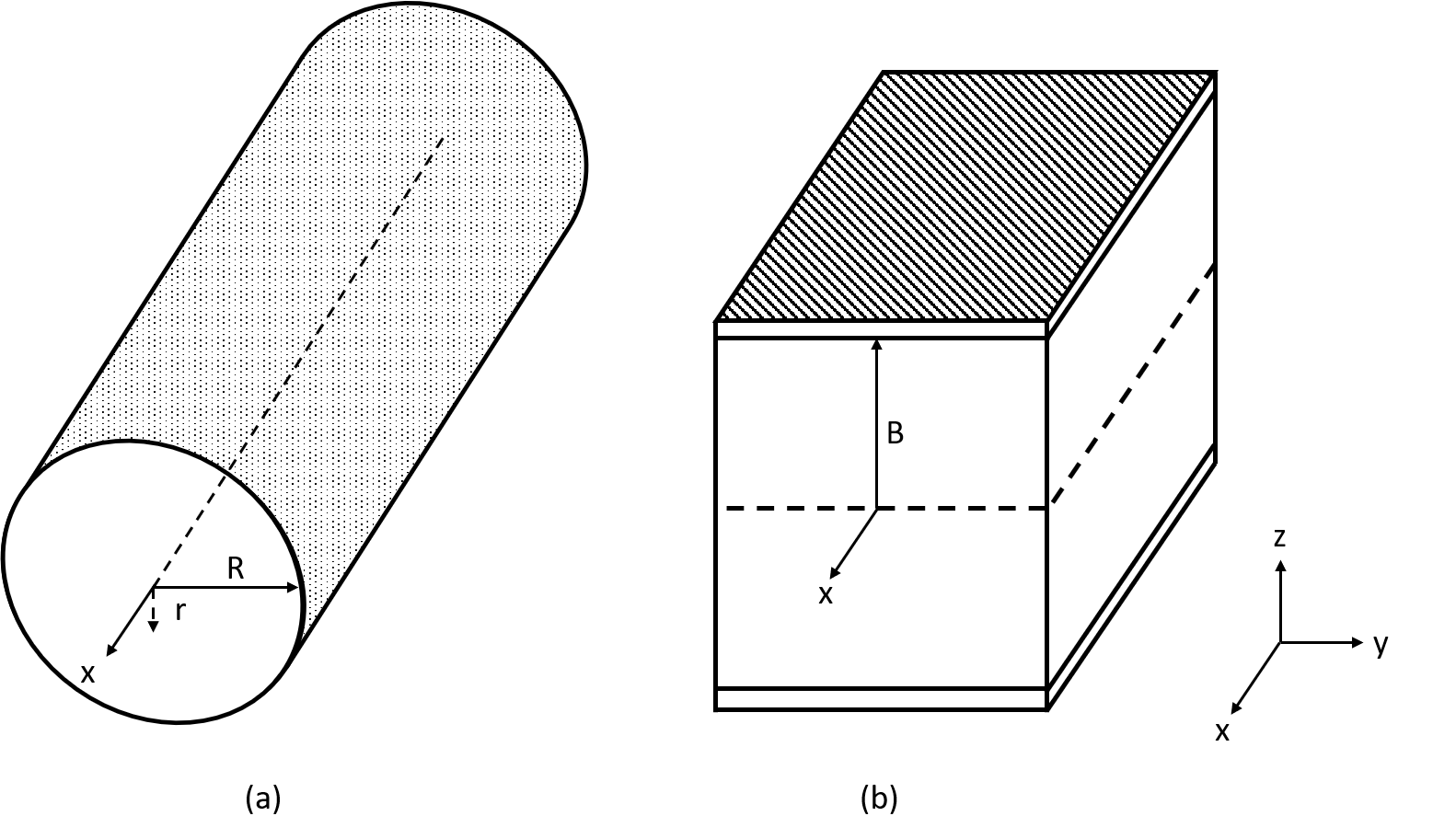}
    \caption{Illustrations of the (a) capillary tube and (b) slit domains. Both figures exhibit a line representing the centrum of the geometry.} 
    \label{fig:geometries}
\end{figure}

The assumptions made here follow those described in the WRMS method \cite{Skelland_67,Sochi_15}. It is assumed that the flow through both geometries is laminar, incompressible, isothermal, steady, pressure-driven, and well developed. The capillary tube is a rigid cylinder that is symmetric about the center, while the slit is symmetric above and below a bisecting line through the system. The slit is  assumed to be infinitely wide such that wall effects are dominated by the confining top and the bottom walls. A representation of each geometry is illustrated in Figure \ref{fig:geometries}. No-slip conditions are also assumed at the wall boundaries for both geometries. Under such conditions, the forces that exist in either system are the shear stresses at the walls of the system and the pressures at the inlets and outlets. A force balance conducted on the capillary tube system reveals that the wall shear stress and change in pressure across the system are related by 
\begin{equation}
\tau_{Rx} = \frac{R \gkD p}{2 L}\;,
\label{eq:tauR}
\end{equation}
where $\tau_{Rx}$ is the magnitude of the shear stress, the $Rx$-subscript follows standard conventions\cite{Bird_Stewart_etal_02} that indicates stress in the $x$-direction acting on an area perpendicular to $R$, $R$ is the radius of the tube, $\gkD p=p_1-p_2$ where $p_1$ and $p_2$ are the average pressures across the inlet and outlet respectively, and $L$ is the length of the tube \cite{Skelland_67,Sochi_15}. A similar force balance for the slit yields
\begin{equation}
\tau_{Bx} = \frac{B\gkD p}{L}\;,
\label{eq:tauB}
\end{equation}
where $B$ is the  half-height of the slit, indicating distance from the centerline of the slit to the wall. These relations can be used to calculate the flow rate within either of these geometries \cite{Sochi_15}, but they hold regardless of the location of the walls \cite{Skelland_67}. Thus, at some arbitrary distance $r$ from the centrum of the capillary tube, the shear stress in the direction of flow can be calculated by \cite{Kim_18}
\begin{equation}
\tau_{rx} = \frac{r\gkD p}{2L}\;,
\label{eq:taur}
\end{equation}
At a distance $z$ from the centrum of the slit formed between parallel plates, the shear stress can be written as\cite{Sochi_15} 
\begin{equation}
\tau_{zx} = \frac{z\gkD p}{L}\;.
\label{eq:taub}
\end{equation}
These relations can be used to calculate the velocity distributions throughout a cross section of each geometry.

For the capillary tube geometry, the magnitude of the velocity at any given distance from the centrum, $v_w(r)$, can be computed using the following analysis,
\begin{align}
v_w(r) 
=\gils 0{v_w(r)} {}{v_w} 
=& \gils Rr {\lrp{\od {v_w}r}}r 
=\gils{\tau_{Rx}}{\tau_{rx}}{\lrp{\od {v_w}r}\lrp{\od r{\tau_w}}}{\tau_w} 
\nonumber \\
&=\gils{\dot{\gkg}_R}{\dot{\gkg}_r}{\dsgiw\lrp{\od r{\tau_w}}\lrp{\od{\tau_w}{\dsgiw}}}{\dsgiw}\;,
\label{eq:vwr}
\end{align}
where $\dot{\gkg}_R$ is the shear rate at the tube wall and $\dot{\gkg}_r$ is the shear rate at the distance $r$. Taking the derivative of \Eqn{taur} with respect to $r$ and substituting into \Eqn{vwr} gives 
\begin{equation}
v_w(r) = \frac{2L}{\gkD p}\gils{\dot{\gkg}_R}{\dot{\gkg}_r}{\dsgiw\lrp{\od{\tau_w}{\dsgiw}}}{\dsgiw}\;,
\label{eq:vwr2}
\end{equation}
which can be solved for a given rheological model by computing d${\tau}/$d${\dot{\gkg}}$ and integrating. 

A similar analysis carried out on the slit geometry yields
\begin{equation}
v_w(z) = \frac{L}{\gkD p}\gils{\dot{\gkg}_B}{\dot{\gkg}_z}{\dsgiw\lrp{\od{\tau_w}{\dsgiw}}}{\dsgiw}\;,
\label{eq:vwb}
\end{equation}
where $\dot{\gkg}_B$ and $\dot{\gkg}_z$ are the shear rate at the slit wall and at distance $z$ respectively. Noting that the integrals in \Eqnstwo{vwr2}{vwb} are the same except for their bounds, a term $I_v$ will be defined as 
\begin{equation}
    I_v = \gils{\dot{\gkg}_1}{\dot{\gkg}_2}{\dsgiw\lrp{\od{\tau_w}{\dsgiw}}}{\dsgiw}\;,
    \label{eq:Iv}
\end{equation}
which can be applied to either geometry. The integrations carried out in the following sections were computed using Wolfram Mathematica \cite{Mathematica}.

\subsection{Cross Model}

Plugging \Eqn{Cross} into \Eqn{tauiw} and taking the derivative of the magnitude of the shear stress in the direction of the flow with respect to $\dsgiw$ gives
\begin{equation}
\od{\tau_w}{\dsgiw} = \hsmu_\infty + \frac{\hsmu_0-\hsmu_\infty}{1+\lrp{m\dsgiw}^n} - \frac{n\lrp{\hsmu_0-\hsmu_\infty}\lrp{m\dsgiw}^n}{\lrb{1+\lrp{m\dsgiw}^n}^2}\;.
\label{eq:odtgCross}
\end{equation}
Substituting \Eqn{odtgCross} into \Eqn{Iv} and integrating gives
\begin{align}
I_v =& \frac{\dot{\gkg}_1^2\lrb{\lrp{\hsmu_0-\hsmu_\infty} g_1\; _2F_1\lrp{1,\frac 2n; \frac{n+2}{n};-f_1 } -2\lrp{\hsmu_0-\hsmu_\infty}-\hsmu_\infty g_1 } }{2g_1 }
\nonumber \\
&- \frac{\dot{\gkg}_2^2\lrb{\lrp{\hsmu_0-\hsmu_\infty} g_2\; _2F_1\lrp{1,\frac 2n; \frac{n+2}{n};-f_2 } -2\lrp{\hsmu_0-\hsmu_\infty}-\hsmu_\infty g_2 } }{2g_2 }\;, 
\label{eq:Iv_Cross}
\end{align}
where $f_1=\lrp{m\dot{\gkg}_1}^n$, $f_2=\lrp{m\dot{\gkg}_2}^n$, $g_1 = 1+f_1$, and $g_2 = 1+f_2$, following the conventions laid out by Sochi (2015) \cite{Sochi_15}, and ${_2F_1}$ is the hypergeometric function. By substituting \Eqn{Iv_Cross} into \Eqnstwo{vwr2}{vwb}, analytical solutions are completed for velocity in a capillary tube or slits between parallel plates. These analytical solutions are not closed, as the shear rate at the wall and distance from the centrum need to be computed numerically, as described previously. 

\subsection{Carreau Model}

For Carreau model fluids, the same procedure is followed as for the Cross model. \Eqn{Carreau} is substituted into \Eqn{tauiw}, and the derivative of the magnitude of the shear stress in the direction of the flow with respect to $\dsgiw$ is computed, giving 
\begin{equation}
\od{\tau_w}{\dsgiw} = \hsmu_{\infty} + \frac{\lrp{\hsmu_0-\hsmu_{\infty}}\lrb{1-\lrp{n'-1}\lrp{m\dsgiw}^2} }{\lrb{1+\lrp{m\dsgiw}^2 }^{\frac{n'+2}{2}} }\;,
\label{eq:odtgCarreau}
\end{equation}
where $n'=1-n$, following a similar convention to what is used in Sochi (2015)\cite{Sochi_15}. Again, substituting \Eqn{odtgCarreau} into \Eqn{Iv} and integrating gives
\begin{align}
I_v = &\frac{2\lrp{\hsmu_0-\hsmu_{\infty}}\lrb{1+\lrp{m\dot{\gkg}_2}^2}^{-n'/2}\lrb{\lrp{m\dot{\gkg}_2}^2\lrp{n'-1}+1}+\hsmu_{\infty}\lrp{m\dot{\gkg}_2}^2\lrp{n'-2} }{2 m^2\lrp{n'-2}} 
\nonumber \\ 
&-\frac{2\lrp{\hsmu_0-\hsmu_{\infty}}\lrb{1+\lrp{m\dot{\gkg}_1}^2}^{-n'/2}\lrb{\lrp{m\dot{\gkg}_1}^2\lrp{n'-1}+1}+\hsmu_{\infty}\lrp{m\dot{\gkg}_1}^2\lrp{n'-2} }{2 m^2\lrp{n'-2}} \;,
\label{eq:Iv_Carreau}
\end{align}
which completes the calculation for Carreau model fluids when \Eqn{Iv_Carreau} is substituted into \Eqnstwo{vwr2}{vwb}.

\subsection{Computing Velocities}

The analytical solutions for velocity are not in a closed form, as they rely on computation of shear rates that do not have an analytical solution with respect to shear stress. The shear rates, $\dot{\gkg}_r$ and $\dot{\gkg}_R$ for the capillary tube case and $\dot{\gkg}_z$ and $\dot{\gkg}_B$ for the slit, are calculated by first computing the shear stress using \Eqnstwo{taur}{taub} for a given $\gkD p$. These shear stresses are used with the generalized form of Newton's law of viscosity, becoming  
\begin{equation}
    \tau_{rx} = \hsmu_w\lrp{\dsgiw}\dsgiw\;,
    \label{eq:tauw}
\end{equation}
for the capillary tube case\cite{Bird_Stewart_etal_02,Sochi_15,Kim_18}. For the parallel plates the stress on the left-hand-side of \Eqn{tauw} is $\tau_{zx}$. To compute $\dsgiw$ from \Eqn{tauw}, the corresponding rheological model from \Eqnstwo{Cross}{Carreau} is substituted for $\hsmu_w\lrp{\dsgiw}$, and a root-finding algorithm is used to compute the corresponding shear rate. The shear rates computed in this way are then substituted into \Eqnstwo{Iv_Cross}{Iv_Carreau} depending on the rheology, and then $I_v$ is substituted into \Eqnstwo{vwr2}{vwb} depending on the geometry. While the velocity calculations rely on numerical root-finding, this computation can be done to a specified precision, which should be based on the precision of the algorithm used to find $\dsgiw$. In this work, the \texttt{vpasolve} algorithm built into Matlab \cite{Matlab_24} was used to find $\dsgiw$ to 32-digits of precision.

\section{Advective-Dominated Species Transport}

The analytical solutions for species transport of a passive scalar in capillary tubes and slits when transport is dominated by advection are discussed here. The point where advection or molecular diffusion dominate species transport is determined by the Peclet number, Pe, defined by\cite{Chaplain_Allain_etal_98,Boschan_Ippolito_etal_08,An_Sahimi_etal_22}
\begin{equation}
\text{Pe} = \frac{v^{\overline{w}}L}{\hsD_{m,i}}\;,
\end{equation}
where the superscript with the overbar indicates a density weighted average over the $w$-domain of velocity magnitude\cite{Gray_Miller_14}, and $\hsD_{m,i}$ is the molecular diffusion coefficient of species $i$. Here, the concentration of a dilute species $i$ in the $w$ phase is modeled, defined as $C_{iw}$ at the microscale. It is also typical to model the average concentration of a species over a cross-section of the geometry. In a capillary tube, at some location $x$ along the tube, the average concentration at a cross section can be computed from 
\begin{equation}
\ao{C_{iw}(x,t)}_A = \frac{1}{\pi R^2}\ilims{A}{}\Cisw(x,t)\dif A = \frac{2}{R^2}\ilims{0}{R}\Cisw(x,t) r\dif r\;,
\label{eq:CiACap}
\end{equation}
where $\ao{}_A$ is an averaging operator taken over the cross-sectional area of the system perpendicular to the direction of flow. For the slit geometry, this average over a cross-section becomes
\begin{equation}
\ao{C_{iw}(x,t)}_A = \frac{1}{2BW}\ilims{A}{}\Cisw(x,t)\dif A = \frac{1}{B}\ilims{0}{B}\Cisw(x,t) \dif z\;,
\label{eq:CiASlit}
\end{equation}
where averaging has reduced the dimensionality to one dimension in space, $x$, and time $t$---as explicitly denoted.

The initial condition of each geometry is that there is no species present beyond the inlet of the geometry, while there is a constant concentration $C_0$ of the species at the inlet. This may be formulated as
\begin{equation}
\Cisw\lrp{x,a,0} =
\begin{cases}
C_0 & x \leq 0 \\
0 & x > 0\;,
\end{cases}
\label{eq:initial_condition}
\end{equation}
where $a$ indicates the distance from the centrum, corresponding to $r$ for the capillary tube and $z$ for the slit, which is introduced for generality and to avoid repetition.

At a high Pe, the effects of diffusion may be neglected, and the concentration of solute at distance $a$ and time $t$ can be formulated in terms of the Heaviside function $H$ as \cite{Taylor_53} 
\begin{equation}
\Cisw\lrp{x,a,t} = C_0 H\lrp{x^*}\;,
\label{eq:CiswXYT}
\end{equation}
where $x^*$ is defined as
\begin{equation}
    x^* = v_w\lrp{a}t - x\;.
\end{equation}
Substituting \Eqn{CiswXYT} into \Eqn{CiACap} is how Taylor computed average concentrations for this case \cite{Taylor_53}; however, a closed form of velocity in terms of $a$ was known for Newtonian fluids. To generalize this approach to cases where a closed-form for velocity is not accessible, the quantity $a^*$ is introduced here, which is the distance from the centrum at which 
\begin{equation}
v_w\lrp{a^*}t = x\;,
\label{eq:vy*}
\end{equation}
or where $x^*$ is zero. It can be seen by substituting \Eqn{CiswXYT} into \Eqn{CiACap} that
\begin{equation}
    \ao{C_{iw}}_A = \frac{2}{R^2}\ilims{0}{a^*}C_0 r\dif r\;,
\end{equation}
and thus, after normalizing by $C_0$, for the capillary tube case
\begin{equation}
\ao{\Ciswo}_A\lrp{x,t} =
\begin{cases}
1 & x \leq 0 \\
\lrp{a^*}^2/R^2  & 0 < x \leq v_w\lrp{0}t \\
0 & x> v_w\lrp{0}t
\end{cases}\;,
\label{eq:CiA_analyticCap}
\end{equation}
where $\Ciswo = \Cisw/C_0$. For the slit case, \Eqn{CiswXYT} is substituted into \Eqn{CiASlit} and normalized by $C_0$, giving
\begin{equation}
\ao{\Ciswo}_A\lrp{x,t} =
\begin{cases}
1 & x \leq 0 \\
a^*/B & 0 < x \leq v_w\lrp{0}t \\
0 & x> v_w\lrp{0}t
\end{cases}\;.
\label{eq:CiA_analyticSlit}
\end{equation}
In the case that a closed-form solution for velocity as a function of $a$ is known, $a^*$ is computed from \Eqn{vy*} directly. In the case of Cross and Carreau model fluids here, where an open-form solution for velocity is known, $a^*$ is computed numerically. It should be noted that, when using the equation for $v_w\lrp{a}$ during Newtonian fluid flow in a capillary tube, \Eqn{CiA_analyticCap} simplifies to Taylor's original solution \cite{Taylor_53} for these initial and boundary conditions. 

\section{Enhanced Molecular Diffusion}
\label{sec:enhancedDm}

Enhanced molecular diffusion occurs at intermediate Pe where advection is the dominant transport phenomena in the longitudinal direction, and molecular diffusion is dominant in the directions transverse to flow. The derivation of the general analytical solution for enhanced molecular diffusion during dilute species transport follows Taylor's analysis \cite{Taylor_53} by first normalizing the species conservation of mass equation. Then, the normalized conservation equations undergo a change of variables to a moving frame of reference. Finally, the conservation of mass equation is solved for $\Cisw$ in the moving frame of reference for the general case, without a closed-form solution to velocity being necessary. After the conservation of mass equation is solved, the enhanced molecular diffusion coefficient can be computed from the general solution for $\Cisw$.

\subsection{Conservation of Mass Equations}

The general conservation of mass equation, written in Cartesian coordinates for an inert species, is given as
\begin{equation}
    \pdt{\Cisw} + \del\cdot\lrp{\Cisw\visw} = 0\;,
    \label{eq:con_mass0}
\end{equation}
where $\visw$, the species velocity, which can be treated by breaking it into the phase velocity and a deviation velocity $\uiw$, as in 
\begin{equation}
\visw = \viw + \uiw\;.
\label{eq:visw}
\end{equation}
The deviation velocity is then removed using the following Fickian assumption 
\begin{equation}
\Cisw\uiw = -\hsD_{m,i}\del\Cisw\;.
\label{eq:Fickian}
\end{equation}
Substituting \Eqnstwo{visw}{Fickian} into \Eqn{con_mass0}, assuming steady incompressible flow, and a passive scalar unaffected by composition, the conservation of mass equation becomes the advection-diffusion equation given by
\begin{equation}
 \pdt{\Cisw} + \viw\cdot\del\Cisw - \hsD_{m,i}\del^2\Cisw = 0\;.
 \label{eq:con_mass}
\end{equation}
With flow confined to only the $x$-direction, the dot product between velocity and the concentration gradient becomes
\begin{equation}
\viw\cdot\del\Cisw = v_{w}\pd{\Cisw}{x}\;.
\label{eq:vdelCisw}
\end{equation}

Molecular diffusion is assumed to only be important in the $z$ direction, where it is the dominant transport mechanisms, thus the Laplacian term on the left-hand side of \Eqn{con_mass} can be re-written in a form that depends on the geometry. For the capillary tube represented in cylindrical coordinates and assuming symmetry in the cross-sectional dimensions, the Laplacian term becomes
\begin{equation}
    \del^2\Cisw = \pdn{2}{\Cisw}{r} + \frac{1}{r}\pd{\Cisw}{r}\;.
\end{equation}
For the slit geometry, the Laplacian is 
\begin{equation}
\del^2\Cisw = \pdn{2}{\Cisw}{z}\;.
\end{equation}
For the capillary tube, \Eqn{con_mass} is normalized by the radius of the tube, becoming
\begin{equation}
    \frac{R^2}{\hsD_{m,i}}\pdt{\Cisw} + \frac{R^2}{\hsD_{m,i}}v_w\pd{\Cisw}{x} - \pdn{2}{\Cisw}{z_R} - \frac{1}{z_R}\pd{\Cisw}{z_R} = 0\;,
    \label{eq:conmassR}
\end{equation}
where 
\begin{equation}
    z_R = \frac{r}{R}\;.
\end{equation}
To generate the conservation of mass equation for the slit geometry, \Eqn{con_mass} is normalized by $B$, becoming
\begin{equation}
    \frac{B^2}{\hsD_{m,i}}\pdt{\Cisw} + \frac{B^2}{\hsD_{m,i}}v_w\pd{\Cisw}{x} - \pdn{2}{\Cisw}{z_B} = 0\;,
\end{equation}
where 
\begin{equation}
    z_B = \frac{z}{B}\;.
\end{equation}

\subsection{Change in Variables}

The next step in the Taylor-like analysis carried out here is to change to a Lagrangian frame of reference $\tilde{x}$ that moves with the average macroscale velocity of the fluid $v^{\overline{w}}$, defined by
\begin{equation}
    \tilde{x} = x - v^{\overline{w}}t\;.
\end{equation}
This change in variables is carried out for both geometries, and the conversion is the same, although the calculation will explicitly be shown for the capillary tube case here. First, note that the total derivative for $\Cisw$ with respect to $t$ as a function of $\tilde{x}$ is
\begin{equation}
\od{\Cisw}{t} =  \pdt{\Cisw} + \pd{\Cisw}{\tilde{x}}\od{\tilde{x}}{t}\;.
\label{eq:odCt}
\end{equation}
where the derivative of $\tilde{x}$ is
\begin{equation}
\od{\tilde{x}}{t} = v_w - v^{\overline{w}}\;.
\label{eq:odx1t}
\end{equation}
It may be seen that the total derivative could have also been written in-terms of $x$ as
\begin{equation}
\od{\Cisw}{t} = \pdt{\Cisw} + v_w\pd{\Cisw}{x}\;,
\label{eq:odCt2}
\end{equation}
due to velocity being zero in the $z_R$ direction. Substituting \Eqnstwo{odCt}{odx1t} into \Eqn{odCt2} and rearranging gives
\begin{equation}
v_w\pd{\Cisw}{x} = \lrp{v_w-v^{\overline{w}}}\pd{\Cisw}{\tilde{x}}\;,
\label{eq:vpdCx}
\end{equation}
which, when substituted into \Eqn{conmassR}, completes the change in variables. Thus, for the capillary tubes case, the conservation of mass equation becomes
\begin{equation}
    \frac{R^2}{\hsD_{m,i}}\pdt{\Cisw} + \frac{R^2}{\hsD_{m,i}}\lrp{v_w-v^{\overline{w}}}\pd{\Cisw}{\tilde{x}} - \pdn{2}{\Cisw}{z_R} - \frac{1}{z_R}\pd{\Cisw}{z_R} = 0\;.
    \label{eq:con_massCap}
\end{equation}
Similarly, the slit conservation of mass equation becomes
\begin{equation}
    \frac{B^2}{\hsD_{m,i}}\pdt{\Cisw} + \frac{B^2}{\hsD_{m,i}}\lrp{v_w-v^{\overline{w}}}\pd{\Cisw}{\tilde{x}} - \pdn{2}{\Cisw}{z_B} = 0\;.
    \label{eq:con_massSlit}
\end{equation}

\subsection{Partial Differential Equation Solution}

It has been observed by Taylor \cite{Taylor_53} that, under the conditions of enhanced molecular diffusion, the change in $\Cisw$ with respect to time will be insignificant in the moving frame of reference and thus, after normalizing by $C_0$, \Eqnstwo{con_massCap}{con_massSlit} become
\begin{equation}
    \frac{R^2}{\hsD_{m,i}}\lrp{v_w-v^{\overline{w}}}\pd{\Ciswo}{\tilde{x}} - \pdn{2}{\Ciswo}{z_R} - \frac{1}{z_R}\pd{\Ciswo}{z_R} = 0\;,
    \label{eq:con_massCap2}
\end{equation}
and 
\begin{equation}
    \frac{B^2}{\hsD_{m,i}}\lrp{v_w-v^{\overline{w}}}\pd{\Ciswo}{\tilde{x}} - \pdn{2}{\Ciswo}{z_B} = 0\;,
    \label{eq:con_massSlit2}
\end{equation}
respectively, where $\Ciswo=\Cisw/C_0$. To solve these partial differential equations, a Taylor-like analysis is carried out where, for the capillary tube case, it is posited that
\begin{equation}
    \Ciswo\lrp{\tilde{x},z_R} = g(\tilde{x}) + \pd{\Ciswo}{\tilde{x}}f(z_R)\;,
    \label{eq:Ciswgf}
\end{equation}
where $g(\tilde{x})$ and $f(z_R)$ are independent of one another, and $\partial\Ciswo/\partial \tilde{x}$ is independent of $z_R$ \cite{Taylor_53}. 

From the posited form given by \Eqn{Ciswgf}, it can be observed that the dependence on $z_R$ is contained in $f(z_R)$. We seek to solve for this function and must have consistency with \Eqn{Ciswgf}, which contains dependency on $z_R$ in the last two terms. Since $\pdil{\Ciswo}{\tilde x}$ is not a function of $z_R$, it only contributes a constant in the solution for $f(z_R)$, which allows \Eqn{con_massCap2}, after multiplication by $z_R$, to be written as
\begin{equation}
    \frac{R^2}{\hsD_{m,i}}\lrp{v_w-v^{\overline{w}}}z_R - z_Rf'' - f' = 0\;.
    \label{eq:fprime}
\end{equation}
Integrating \Eqn{fprime} with respect to $z_R$, the equation becomes
\begin{equation}
    \frac{R^2}{\hsD_{m,i}}\ilims{0}{z_R}\lrp{v_w-v^{\overline{w}}}z_R\dif{z_R} - \ilims{0}{z_R}z_Rf''\dif{z_R} - \ilims{0}{z_R}f'\dif{z_R} = 0\;.
    \label{eq:fprime2}
\end{equation}
Using integration by parts, it can be seen that 
\begin{equation}
    \ilims{0}{z_R}z_Rf''\dif{z_R} = z_Rf' - \ilims{0}{z_R}f'\dif{z_R}\;,
\end{equation}
and thus, \Eqn{fprime2} simplifies to
\begin{equation}
    \frac{R^2}{\hsD_{m,i}}\ilims{0}{z_R}\lrp{v_w-v^{\overline{w}}}z_R\dif{z_R} - z_Rf' = 0\;.
\end{equation}
Re-arranging and integrating again with respect to $z_R$ gives,
\begin{equation}
    f(z_R) = \frac{R^2}{\hsD_{m,i}}\ilims{0}{z_R}\lrb{\frac{1}{z_R}\ilims{0}{z_R}\lrp{v_w-v^{\overline{w}}}z_R\dif{z_R}}\dif{z_R} + f(0)\;.
    \label{eq:fzr}
\end{equation}
To remove $g(\tilde{x})$ and $f(0)$ from the final solution, \Eqn{fzr} is substituted into \Eqn{Ciswgf} and evaluated at $z_R=0$, which when rearranged yields 
\begin{equation}
    g(\tilde{x}) = \Ciswo\lrp{\tilde{x},0} - \pd{\Ciswo}{\tilde{x}}f(0)\;.
    \label{eq:gx1}
\end{equation}
Plugging \Eqnstwo{fzr}{gx1} into \Eqn{Ciswgf} gives
\begin{equation}
    \Ciswo\lrp{\tilde{x},z_R} = \Ciswo\lrp{\tilde{x},0} + \frac{R^2}{\hsD_{m,i}}\pd{\Ciswo}{\tilde{x}}\ilims{0}{z_R}\lrb{\frac{1}{z_R}\ilims{0}{z_R}\lrp{v_w-v^{\overline{w}}}z_R\dif{z_R}}\dif{z_R}\;,
    \label{eq:Ciswx1zR}
\end{equation}
which is a solution to \Eqn{con_massCap2} and is a more general form of Taylor's solution, without the need to have a closed-form solution for velocity with respect to $z_R$. This can be seen by evaluating \Eqn{Ciswx1zR} for a Newtonian fluid, which simplifies to Taylor's solution.

Deriving a Taylor-like function for the slit case can be carried out in a similar  way as for the capillary tube by writing \Eqn{Ciswgf} in terms of $z_B$. Using \Eqn{con_massSlit2} and following similar steps to those used above yields
\begin{equation}
    \Ciswo\lrp{\tilde{x},z_B} = \Ciswo\lrp{\tilde{x},0} + \frac{B^2}{\hsD_{m,i}}\pd{\Ciswo}{\tilde{x}}\ilims{0}{z_B}\ilims{0}{z_B}\lrp{v_w-v^{\overline{w}}}\dif{z_B}\dif{z_B}\;.
    \label{eq:Ciswx1zB}
\end{equation}

\subsection{Enhanced Diffusion Coefficient}

To formulate the enhanced molecular diffusion coefficient, Taylor \cite{Taylor_53} computed the mass transfer rate between cross-sections; however, a different approach will be used here that follows recent developments in the porous media literature. Specifically, some of the insights generated by treating species transport using thermodynamically constrained averaging theory (TCAT) will be utilized \cite{Sciume_Shelton_etal_12,Sciume_Shelton_etal_13,Gray_Miller_14,Weigand_Schultz_etal_18,Miller_Gray_etal_21}. While a detailed treatment of TCAT is outside of the scope of this work, there are many applications of the theory in the literature, including to single fluid flow \cite{Gray_Miller_06,Miller_Gray_08,Gray_Miller_09,Gray_Miller_09c,Gray_Miller_14}, two fluid flow \cite{Jackson_Miller_etal_09,Gray_Miller_14,Gray_Dye_etal_15,Rybak_Gray_etal_15}, sediment transport \cite{Miller_Gray_etal_19}, and non-Newtonian fluid flow and transport\cite{Bowers_Miller_21,Bowers_Miller_23,Bowers_Miller_25}. An aspect of TCAT that is of use here is that every averaged scale, called macroscale, quantity is precisely defined as an average of defined fluid continuum scale, called microscale, quantities \cite{Gray_Miller_14}. Thus, models that are developed at the macroscale can be either validated or informed by microscale observations.

One aspect of TCAT that needs to be explicitly defined here is the averaging convention. A general average is formulated in TCAT using
\begin{equation}
\big\langle{F}\big\rangle_{\Dm\gkb,\Dm\gkg, W} = \gaint F{\Dm \gkb}{\Dm \gkg}{W}\;,
\label{eq:aop}
\end{equation}
where $F$ is an arbitrary microscale quantity, $W$ is the weighting function, and $\Dm \gkb$ and $\Dm \gkg$ are averaging domains \cite{Gray_Miller_14}. There are several kinds of averages that are utilized in TCAT, but they are all specific instances of \Eqn{aop}. 

It has been shown that, during linear Fickian dispersion of a dilute passive scalar, hydrodynamic dispersion may be defined as averages of microscale quantities by \cite{Gray_Miller_14,Weigand_Schultz_etal_18} 
\begin{equation}
    \aop{\riw\gko_{iw}}{w}{w}{}\vec{u}^{\dol{iw}} = -\htD^{iw}\vdot\del\aop{\riw\gko_{iw}}{w}{w}{}\;,
    \label{eq:macroFickian}
\end{equation}
where $\riw$ is the density of the solvent, $\gko_{iw}$ is the mass fraction of species $i$, $\vec{u}^{\dol{iw}}$ is the macroscale deviation velocity, defined explicitly as
\begin{equation}
    \vec{u}^{\dol{iw}} = \aop{\visw}{w}{w}{\riw\gko_{iw}}-\aop{\viw}{w}{w}{\riw}\;,
    \label{eq:uasw}
\end{equation}
and $\htD^{iw}$ is the macroscale hydrodynamic dispersion tensor for species $i$ \cite{Gray_Miller_14,Weigand_Schultz_etal_18}. Substituting \Eqn{uasw} into \Eqn{macroFickian} and recalling the definition presented in \Eqn{aop} yields
\begin{equation}
    \aop{\riw\gko_{iw}\visw}{w}{w}{}-\aop{\riw\gko_{iw}}{w}{w}{}\aop{\viw}{w}{w}{\riw}=-\htD^{iw}\vdot\del\aop{\riw\gko_{iw}}{w}{w}{}\;.
    \label{eq:macroFickian2}
\end{equation}
The concentration of the dilute species is defined as  $\Cisw=\riw\gko_{iw}$ and, because the fluid is assumed to be of constant density, the density-weighted velocity average is equivalent to a volume average velocity. Thus, \Eqn{macroFickian2} can be written for the one-dimensional macroscale case considered here as 
\begin{equation}
    \aop{\Cisw v_{iw}}{w}{w}{}-\aop{\Cisw}{w}{w}{}v^{\overline{w}}=-\hsD^{iw}\pd{}{x}\aop{\Cisw}{w}{w}{}\;.
    \label{eq:macroFickian3}
\end{equation}
Substituting \Eqnstwo{visw}{Fickian} into \Eqn{macroFickian3} to remove the species velocity within the averaging operator yields
\begin{equation}
    \aop{\Cisw v_{w}}{w}{w}{}-\hsD_{m,i}\aop{\pd{\Cisw}{x}}{w}{w}{}-\aop{\Cisw}{w}{w}{}v^{\overline{w}}=-\hsD^{iw}\pd{}{x}\aop{\Cisw}{w}{w}{}\;.
    \label{eq:macroFickian4}
\end{equation}
Using a gradient averaging theorem [Eqn (B.13) in Gray and Miller (2014)\cite{Gray_Miller_14}] and rearranging gives
\begin{equation}
    \aop{\Cisw \lrp{v_{w}-v^{\overline{w}}}}{w}{w}{}-\hsD_{m,i}\pd{}{x}\aop{\Cisw}{w}{w}{}=-\hsD^{iw}\pd{}{x}\aop{\Cisw}{w}{w}{}\;.
    \label{eq:macroFickian5}
\end{equation}
The form of the above equation was chosen because the first term on the left-hand-side can be shown to be equivalent to Taylor's mass transfer rate between cross-sections at location $\tilde{x}$ [Eqn (22) in Taylor (1953)\cite{Taylor_53}], maintaining consistency with the literature.

To complete the calculation, \Eqn{macroFickian5} is divided by $C_0$, and the solution for $\Ciswo$ presented for each geometry is substituted into the resulting equation. For the capillary tube case, this results in \Eqn{Ciswx1zR} being substituted, giving
\begin{align}
    \aop{\Ciswo\lrp{\tilde{x},0}\lrp{v_{w}-v^{\overline{w}}}}{w}{w}{}&+\aop{\frac{R^2}{\hsD_{m,i}}\pd{\Ciswo}{\tilde{x}}I_{R1}\lrp{z_R}\lrp{v_{w}-v^{\overline{w}}} }{w}{w}{}
    \nonumber \\
    &-\hsD_{m,i}\pd{}{x}\aop{\Ciswo}{w}{w}{}=-\hsD^{iw}\pd{}{x}\aop{\Ciswo}{w}{w}{}\;.
    \label{eq:aopCiswoCap}
\end{align}
where $I_{R1}$ is defined for clarity as the integral
\begin{equation}
    I_{R1}\lrp{z_R} = \ilims{0}{z_R}\frac{1}{z_R}I_{R2}\lrp{z_R}\dif{z_R}\;,
    \label{eq:IR1}
\end{equation}
and $I_{R2}$ is the integral
\begin{equation}
    I_{R2}\lrp{z_R} = \ilims{0}{z_R}\lrp{v_w-v^{\overline{w}}}z_R\dif{z_R}\;.
    \label{eq:IR2}
\end{equation}
Because $\Ciswo\lrp{\tilde{x},0}$ is spatially independent of $\lrp{v_w-v^w}$, the first average in \Eqn{aopCiswoCap} becomes the product of averages of $\Ciswo\lrp{\tilde{x},0}$ and $\lrp{v_w-v^w}$, which becomes zero after averaging. Similarly, converting the second average in \Eqn{aopCiswoCap} into the product of averages of independent quantities, \Eqn{aopCiswoCap} becomes
\begin{align}
    \frac{R^2}{\hsD_{m,i}}
    &\aop{\pd{\Ciswo}{\tilde{x}}}{w}{w}{}
    \aop{I_{R1}\lrp{z_R}\lrp{v_{w}-v^{\overline{w}}} }{w}{w}{}
    \nonumber \\
    &-\hsD_{m,i}\pd{}{x}\aop{\Ciswo}{w}{w}{}=-\hsD^{iw}\pd{}{x}\aop{\Ciswo}{w}{w}{}\;.
    \label{eq:aopCiswoCap2}
\end{align}

Because the velocity is constant at a fixed position in space and temporal changes in the moving frame of reference are ignored, it follows that
\begin{equation}
    \aop{\pd{\Ciswo}{\tilde{x}}}{w}{w}{} = \pd{}{x}\aop{\Ciswo}{w}{w}{}\;,
\end{equation}
and \Eqn{aopCiswoCap2} becomes
\begin{equation}
    \frac{R^2}{\hsD_{m,i}}
    \aop{I_{R1}\lrp{z_R}\lrp{v_{w}-v^{\overline{w}}} }{w}{w}{}
    -\hsD_{m,i}=-\hsD^{iw}\;.
    \label{eq:aopCiswoCap3}
\end{equation}
The averaging operator can be applied over some representative elementary volume (REV) for enhanced molecular diffusion, which is assumed to take place over an REV length $\ell$ and the cross-section of the geometry perpendicular to flow. In this case, the averaging operator defined in \Eqn{aop} can be re-written for the arbitrary variable $F$ as the intrinsic average over $\Dm w$ as
\begin{equation}
    \aop{F}{w}{w}{} = \frac{\ilims{\ell}{}\ilims{A}{}F\dif{A}\dif{\ell} }{\ilims{\ell}{}\ilims{A}{}\dif{A}\dif{\ell} }\;.
    \label{eq:aop2}
\end{equation}
Noting that all the terms within the averaging operator in \Eqn{aopCiswoCap3} are only dependent on $z_R$, the integrations over $\ell$ cancel out of \Eqn{aop2}, and it can be shown that
\begin{equation}
\aop{I_{R1}\lrp{z_R}\lrp{v_{w}-v^{\overline{w}}} }{w}{w}{} = 
2\ilims{0}{1}I_{R1}\lrp{z_R}\lrp{v_w-v^{\overline{w}}}z_R\dif{z_R}\;,
\end{equation}
thus the enhanced molecular diffusion coefficient can be computed from
\begin{equation}
\hsD^{iw}=-\frac{2R^2}{\hsD_{m,i}}\ilims{0}{1}I_{R1}\lrp{z_R}\lrp{v_w-v^{\overline{w}}}z_R\dif{z_R}
+\hsD_{m,i}\;.
\label{eq:hsDwCapF}
\end{equation}
Using the closed-form solution for $v_w$ during Newtonian flow in a capillary tube, \Eqn{hsDwCapF} simplifies to Taylor's solution, [Eqn (25) in Taylor (1953)\cite{Taylor_53}]. In the case that an open-form solution exists, as for Cross and Carreau model fluids, the velocities at a set of data points can be tabulated and used to integrate Eqns (\ref{eq:IR1}), (\ref{eq:IR2}), and (\ref{eq:hsDwCapF}) numerically to arbitrary accuracy. Note that the solution presented in \Eqn{hsDwCapF} has not been derived for Cross and Carreau model fluids specifically, and can be used as a general solution for other fluid rheological models.

Computing the enhanced molecular diffusion coefficient for a slit follows the same procedure as was carried out for the capillary tube case. Substituting \Eqn{Ciswx1zB} into \Eqn{macroFickian5} after dividing by $C_0$ yields
\begin{align}
    \aop{\Ciswo\lrp{\tilde{x},0}\lrp{v_{w}-v^{\overline{w}}}}{w}{w}{}&+\aop{\frac{B^2}{\hsD_{m,i}}\pd{\Ciswo}{\tilde{x}}I_B\lrp{z_B}\lrp{v_{w}-v^{\overline{w}}} }{w}{w}{}
    \nonumber \\
    &-\hsD_{m,i}\pd{}{x}\aop{\Ciswo}{w}{w}{}=-\hsD^{iw}\pd{}{x}\aop{\Ciswo}{w}{w}{}\;,
    \label{eq:aopCiswoSlit}
\end{align}
where $I_B$ has been introduced for clarity and is given by 
\begin{equation}
    I_B\lrp{z_B} = \ilims{0}{z_B}\ilims{0}{z_B}\lrp{v_w-v^{\overline{w}}}\dif{z_B}\dif{z_B}\;.
    \label{eq:IB}
\end{equation}
Carrying out the same steps as was done for the tube case, it can be seen that \Eqn{aopCiswoSlit} simplifies to
\begin{equation}
\frac{B^2}{\hsD_{m,i}}
\aop{I_{B}\lrp{z_B}\lrp{v_{w}-v^{\overline{w}}} }{w}{w}{}
-\hsD_{m,i}=-\hsD^{iw}\;.
\label{eq:aopCiswoSlit2}
\end{equation}
Noting that all terms within the averaging operator in \Eqn{aopCiswoSlit2} are dependent only on $z_B$, and again referring to \Eqn{aop2}, it can be shown that 
\begin{equation}
\aop{I_{B}\lrp{z_B}\lrp{v_{w}-v^{\overline{w}}} }{w}{w}{}
= \ilims{0}{1}I_B\lrp{z_B}\lrp{v_w-v^{\overline{w}}}\dif{z_B}\;,
\end{equation}
and thus, the enhanced molecular diffusion coefficient may be computed from 
\begin{equation}
    \hsD^{iw} = -\frac{B^2}{\hsD_{m,i}}\ilims{0}{1}I_B\lrp{z_B}\lrp{v_w-v^{\overline{w}}}\dif{z_B} + \hsD_{m,i}\;.
    \label{eq:hsDwSlitF}
\end{equation}
As described for the capillary tube case, \Eqnstwo{IB}{hsDwSlitF} can be integrated numerically using the open-form solution for velocity. 

\section{Computational Methods}
\label{sec:methods}

The computational methods used here followed a similar procedure to what was done in Bowers and Miller (2025)\cite{Bowers_Miller_25}. Microscale simulations generated a velocity field for the system of interest, and then this velocity field was used to simulate transport of a passive scalar species. The microscale simulation data was used directly to validate the semi-analytical velocity derived in Eqns (\ref{eq:vwr2}) through (\ref{eq:Iv_Carreau}), and the averaged microscale transport simulation data was used to validate \Eqnstwo{CiA_analyticCap}{CiA_analyticSlit}. Averaged microscale transport simulation data were also used to indirectly validate the enhanced molecular diffusion coefficients derived in \Eqnstwo{hsDwCapF}{hsDwSlitF} by comparing the simulation data to a macroscale transport model that utilized the derived coefficients, similar to what was done previously\cite{Weigand_Schultz_etal_18,Weigand_Miller_20,Bowers_Miller_25}. In the following sections the systems that were simulated are described, followed by the microscale simulation methods, macroscale modeling methods, and finally a brief discussion about how data was analyzed to compare simulation and semi-analytical results.

\subsection{System and Fluid Properties}

The capillary tube and parallel plate systems investigated here utilized the same properties described in \S\ref{sec:analyticV} and presented in Figure \ref{fig:geometries}. While the capillary tube can be described solely by $R$ and $L$, simulation of the slit configuration requires $B$, $L$, and $W$ be set, although width does not appear in any of the analytical solutions. The system parameters used are described in Table \ref{tab:sys_param}. The length of the system was selected to be eight times the radius or width, which was found to be required during microscale simulations of transport during GNF flow\cite{Bowers_Miller_25}. The radius and half-height of the systems were selected to be within the same order of magnitude of packed sand, which are a commonly encountered class of porous media\cite{Hauswirth_Bowers_etal_20}. The results generated for these media can be non-dimensionalized by dividing through by $R$ or $B$ and can be applied to other capillary tube and slit media. 

\begin{table}[tbh]
    \centering
    \caption{\label{tab:sys_param} Capillary tube and parallel plate properties}
    \begin{tabular}{lcc}
        Parameter & Capillary Tube & Parallel Plate 
        \\
        \hline
        Length, $L$ (m) & $8.0\times 10^{-3}$ & $8.0\times 10^{-3}$ 
        \\
        Radius/Half-Height, $R$ or $B$ (m) & $5.0\times 10^{-4}$ & $5.0\times 10^{-4}$ 
        \\
        Width, $W$ (m) & N/A & $1.0\times 10^{-3}$ 
        \\
        Number of Cells & 2,852,800 & 1,000,000
        \\
        \hline
    \end{tabular}
\end{table}

Both the Cross and Carreau models have four parameters, and it has become common in theoretical treatments of GNFs to vary each parameter individually\cite{Lashgari_Pralits_etal_12,Tosco_Marchisio_etal_13,Zhang_Prodanovic_etal_19,Alam_Raj_etal_21,Bowers_Miller_21,Bowers_Miller_23,Garg_Prasad_24,Bowers_Miller_25}. While this approach differs from how the parameters of real GNFs typically relate to one another\cite{Casas_Mohedano_etal_00,Escudier_Poole_etal_05,Boschan_Ippolito_etal_08,Hauswirth_Bowers_etal_20}, it makes the fluid mechanical response to each parameter clear, and is thus a useful approach to take. Four Cross and four Carreau model fluids were simulated for this work, with their rheological parameters presented in Table \ref{tab:fluid_param}. Note that each fluid number in Table \ref{tab:fluid_param} refers to two fluids, e.g. both a Cross Fluid 1 and a Carreau Fluid 1 were simulated using the parameters listed as Fluid 1 in the table. The zero and infinite shear viscosities selected are high compared to most real world fluids, but are not outside the realm of what has been observed\cite{Casas_Mohedano_etal_00,Hauswirth_Bowers_etal_20,Wang_22}. It has also been found that the ratio $\hsmu_\infty/\hsmu_0$ has the highest impact on dynamics during flow simulation\cite{Lashgari_Pralits_etal_12}, and the ratios present in Table \ref{tab:fluid_param} are common in the literature, as are the $n$-indices used\cite{Lashgari_Pralits_etal_12,Tosco_Marchisio_etal_13,Zhang_Prodanovic_etal_19,Hauswirth_Bowers_etal_20,Wang_22}. The $m$ parameters were selected to ensure that the full range of GNF dynamic behavior was observed within the laminar flow regime and are not typical; however it has been shown that this parameter only impacts the velocity at which the onset of GNF behavior is observed and not the flow dynamics during the transition from $\hsmu_0$ to $\hsmu_\infty$\cite{Lashgari_Pralits_etal_12,Bowers_Miller_21,Bowers_Miller_23,Bowers_Miller_25}. 

\begin{table}[tbh]
    \centering
    \begin{threeparttable}
    \caption{\label{tab:fluid_param} Cross and Carreau model fluid parameters used}
    \begin{tabular}{lcccc}
        Parameter \hspace{1.5cm} & \hspace{0.5cm} Fluid 1 \hspace{0.5cm} & \hspace{0.5cm} Fluid 2 \hspace{0.5cm} & \hspace{0.5cm} Fluid 3 \hspace{0.5cm} & \hspace{0.5cm} Fluid 4 \hspace{0.5cm} 
        \\
        \hline 
        $\Hat{\mu}_0$ (Pa s) & $1.0\times 10^{2}$ & $1.0\times 10^{3}$ & $1.0\times 10^{2}$ & $1.0\times 10^{2}$ 
        \\
        $\Hat{\mu}_\infty$ (Pa s) & $1.0\times 10^{-1}$ & $1.0\times 10^{-1}$ & $1.0\times 10^{-1}$ & $1.0\times 10^{-1}$  
        \\
        $m$ (s) & $1.0\times 10^{6}$ & $1.0\times 10^{6}$ & $1.0\times 10^{4}$ & $1.0\times 10^{6}$ 
        \\
        $n$ or $n'{}^*$ & $7.0\times 10^{-1}$ & $7.0\times 10^{-1}$ & $7.0\times 10^{-1}$ & $9.0\times 10^{-1}$  
        \\
        $\riw$ (kg/m$^3$) & $1.00\times 10^3$ & $1.00\times 10^3$ & $1.00\times 10^3$ & $1.00\times 10^3$ 
        \\
        \hline
    \end{tabular}
    \begin{tablenotes}
    \item[*] The parameter is $n$ for Cross model fluids and $n'$ for Carreau model fluids.
    \end{tablenotes}
    \end{threeparttable}
\end{table}

\subsection{Microscale Simulation}

To validate the solutions for velocity and species transport derived above, microscale simulations were carried out using OpenFOAM, an open-source computational fluid dynamics package that has been used to simulate GNF flow and transport\cite{Greenshields_18,Zhang_Prodanovic_etal_19,Zheng_19,Hauswirth_Bowers_etal_20,Bowers_Miller_21,Bowers_Miller_23,Amiri_Qajar_etal_24,Bowers_Miller_25}. Microscale simulations followed three steps, starting with mesh generation, followed by velocity field simulation, and finally species transport simulation using the same methods as other recent work\cite{Bowers_Miller_21,Bowers_Miller_23,Bowers_Miller_25}. During species transport simulations, the average concentration at the system outlet was computed using the \texttt{patchAverage} utility built into OpenFOAM, and this averaged quantity was used to validate the analytically derived solution.

Meshes were generated for both the capillary tube and slit configurations by first creating a background mesh using OpenFOAM's \texttt{blockMesh} utility. For the capillary tube case the mesh was then manipulated using the \texttt{searchableCylinder} object and \texttt{snappyHexMesh} utility to resolve the features of the tube, coupled with near-solid refinement which has been found to improve simulation accuracy\cite{Icardi_Boccardo_etal_14,Weigand_Miller_20,Bowers_Miller_21,Bowers_Miller_23,Bowers_Miller_25}. Near solid refinement was not carried out for the slit configuration, as there were no surfaces that needed to be resolved outside the straight walls of the slit. The level of background mesh refinement required to achieve grid-independent results was determined by simulating flow of all the GNFs described in Table \ref{tab:fluid_param} over a range of flow rates such that the full transition between zero- and infinite-shear viscosities were observed at the walls of the system, and then comparing the resulting flow rate to the analytical solution described in Sochi (2015)\cite{Sochi_15}. Once the simulated flow rate exhibited less than 0.5\% error relative to the analytical solution, the system was considered grid-independent. The number of cells required to achieve grid independence are reported in Table \ref{tab:sys_param}.

Flow simulations within the capillary tube and slit systems were driven by applying pressure Dirichlet boundary conditions on the system inlet and outlet to generate a pressure drop. The \texttt{simpleFOAM} solver in OpenFOAM was used to numerically solve the incompressible  conservation of mass equation
\begin{equation}
    \del\cdot\viw=0\;,
\end{equation}
as well as the kinematic conservation of momentum equation defined by
\begin{equation}
    \del\cdot\lrp{\viw\viw}-\del\cdot\tauiw=-\del\piw +\vec S\;,
\end{equation}
where $\vec S$ is a momentum source which was set to zero\cite{Greenshields_18}. At the inlet and outlet of the system velocity was set to have zero gradient Neumann boundary condition. At solid walls the no-slip boundary condition was used for velocity and zero gradient Neumann boundary condition for pressure. For the slit system cyclic boundaries were used for the open sides of the system. The geometric algebraic multigrid (GAMG) numerical method was used, with a relative error tolerance of 1E-8 being set. The \texttt{singleGraph} utility was used to measure velocity across a line with ends located at coordinates $\lrp{4.0\times 10^{-3}, 5.0\times 10^{-3}, 0 }$ and $\lrp{4.0\times 10^{-3}, 5.0\times 10^{-3}, 1.0\times 10^{-3} }$ for both geometries, which cuts through the centrum of the system and lies perpendicular to flow, for comparison to the analytical velocity solution derived in \S\ref{sec:analyticV}. 

Microscale transport simulations were carried out using the \texttt{scalarTransportFoam} utility, which solved the species conservation of mass equation for solute $A$ below,
\begin{equation}
    \pdt{\CiAw}+\del\cdot\lrp{\viw\CiAw}-\del\cdot\lrp{\hsD_{m,A}\del\CiAw}=S_C\;,
\end{equation}
where $S_C$ is a concentration source term\cite{Greenshields_18}. The initial condition used during simulation was the same as was described by \Eqn{initial_condition}. The velocity was taken from the previously described flow simulations. At the inlet a Dirichlet concentration boundary condition was set at 1 mg/L, while a zero gradient Neumann condition was set at the outlet as well as solid walls. A cyclic boundary was set for the open sides of the slit system as well. The molecular diffusion was selected for each simulation to ensure that several Peclet numbers were achieved, given below,
\begin{equation}
    Pe=\lrB{10^{1},10^{2},10^{3},10^{4},10^{5},10^{6}}\;.
\end{equation}
Thus, for each pressure drop simulated, six transport simulations were carried out. This was done to determine at what point advection dominated transport or enhanced molecular diffusion would occur.

All microscale simulations were carried out on 216 cores on the Dogwood computing cluster at the University of North Carolina at Chapel Hill. In total, 50 flow simulations were carried out for each Cross and Carreau model fluid, with an additional 50 carried out for Fluids 4, making for 500 flow simulations. Six microscale transport simulations were carried out for each flow simulation, leading to 3,000 total transport simulations.

\subsection{Macroscale Modeling}

During enhanced molecular diffusion, it is assumed that a linear Fickian assumption may be applied at the averaged system scale, as described in \S\ref{sec:enhancedDm}. To validate the derived enhanced molecular diffusion coefficient, the average effluent concentration computed from the microscale transport simulations was compared to the average concentration computed by solving the one-dimensional advection-dispersion equation, given by
\begin{equation}
    \pdt{}\aop{\CiAw}{w}{w}{}+v^{\overline{w}}\pd{}{x}\aop{\CiAw}{w}{w}{} - \hsD^{Aw}\pdn{2}{}{x}\aop{\CiAw}{w}{w}{}=0\;,
    \label{eq:pdtCaAw}
\end{equation}
where $\hsD^{Aw}$ is computed from \Eqnstwo{hsDwCapF}{hsDwSlitF} depending on the system, and the averaging operator is defined by \Eqn{aop}. To solve \Eqn{pdtCaAw} numerically, a cell-centered finite difference scheme was used to discretize the gradient and Laplacian terms, as well as a backwards Euler method to discretize the time derivative as in past work\cite{Weigand_Schultz_etal_18,Weigand_Miller_20,Bowers_Miller_25}. To more readily observe trends in behavior of the enhanced molecular diffusion with respect to velocity, the longitudinal dispersivity $\hsa^w_L$ was computed from
\begin{equation}
    \hsa^w_L = \frac{1}{v^{\overline{w}}}\lrp{\hsD^{Aw}-\hsD_{m,A}}\;,
    \label{eq:hasawL}
\end{equation}
which has been shown to change depending GNF rheological properties\cite{Bowers_Miller_25}.

\subsection{Computation of Semi-Analytical Solutions}

The analytical solutions of concentration and enhanced molecular diffusion coefficient are not in closed form as these rely on knowledge of the velocities. Thus, each of the analytical solutions provided here require the use of root-finding methods. 

The method laid out in \S~\ref{sec:analyticV} was followed to compute velocities, and the \texttt{vpasolve} numerical solver in Matlab\cite{Matlab_24} was used to find $\dsgiw$. The \texttt{vpasolve} inputs were the symbolic equation and the range of possible shear rates, computed from \Eqn{tauw} assuming either $\hsmu_0$ or $\hsmu_\infty$. Based on the precision of the \texttt{vpasolve} algorithm, $\dsgiw$ was solved to 32-digits of accuracy\cite{Matlab_24}.

To compute the average concentration at the outlet of each geometry during advection dominated flow, $a^*$ was computed numerically. First a velocity profile was computed using the semi-analytical solution at a set of points as described above. Then, the velocity at which $a^*$ occurs was computed from
\begin{equation}
    v_w\lrp{a^*}t = L\;.
\end{equation}
The input velocity profile was then searched to find the bounding velocities that $v_w\lrp{a^*}$ was between, and then a linear interpolation was carried out to compute $a^*$. This value was then used to compute average concentration at the outlet of the system using \Eqnstwo{CiA_analyticCap}{CiA_analyticSlit} depending on the geometry of the system. Similarly to the velocity calculation, the concentration calculation may be carried out to arbitrary accuracy based on the accuracy of $a^*$. Here the number of points of velocity computed for the input velocity profile was 200, which was selected such that $a^*$ changed less than $0.01\%$ relative to its value when doubling the number of points.  

Calculation of the enhanced molecular diffusion coefficient required numerically integrating \Eqnstwo{hsDwCapF}{hsDwSlitF} using an input velocity profile. The velocity profile was computed from the semi-analytical solution as described above. The integrals were then each computed for solute $A$ using the trapezoidal rule, resulting in the numerically computed enhanced molecular diffusion coefficients for each geometry. As with the velocity profile and advection dominated concentration calculations above, $\hsD^{Aw}$ may be computed to arbitrary accuracy, in this case based on the number of velocity points used in the integration. For the enhanced molecular diffusion coefficient, the number of points of velocity used was 200, which was selected such that $\hsD^{Aw}$ changed less than $0.01\%$ relative to its value when doubling the number of points.  

\subsection{Data Analysis}

Validating the semi-analytical solutions present a challenge that is not unique to computational simulation, namely that there are more points of data to analyze than can be succinctly presented. For each of the 500 flow simulations performed, 50 velocity values were output, resulting in 25,0000 possible points of comparison that can be carried out between microscale simulation and the analytical velocity solution. For transport, 200 time steps were output for each of the 3,000 simulations, leading to 600,000 possible points of comparison. 

To compactly present a validation of the analytical solutions versus microscale simulation data, two descriptive graphics more commonly associated with population statistics will be used here, namely a plot of prediction versus observation as well as a box-and-whisker plot. Box-and-whisker plots will show the root-mean-squared-error (RMSE) percent for each pressure drop simulated, as well as the algebraic average of RMSE for each fluid as red markers/lines, and plot ``outliers'' that deviate from the median by more than 1.5 times the inter-quartile range (IQR) as blue markers, as is standard for such plots.  

It is useful to discuss these outliers further, as although they do not have the same meaning that they do in population statistics, they are descriptive of underlying random processes. While two computer simulations carried out with the same inputs and methods would give the same answer, there is an amount of numerical error associated with those simulations that cannot be absolutely predicted. For an analytical simulation, the numerical error will be related to machine precision, while the schemes used in computational fluid dynamics software such as OpenFOAM will exhibit error related to their mesh refinement and numerical methods. Thus, data that lay outside the IQR in this case will define the numerical fidelity of simulations with respect to the physics being modeled.

\section{Results}

Each of the analytically derived models were validated using averaged microscale simulation data. The pressure drops that were simulated are tabulated in the supplementary material in Table SI. The pressure drops listed in that table were used to identify simulations in the below sections. The semi-analytical velocity solution validations are presented first, as these are required to compute the two transport models derived. The advection dominated species transport and enhanced molecular diffusion models were then validated respectively. The capillary tube and slit configurations are presented together for each model.

\subsection{Semi-Analytical Velocity Solutions}

\begin{figure}[t]
   \centering
    \includegraphics[width=1\linewidth]{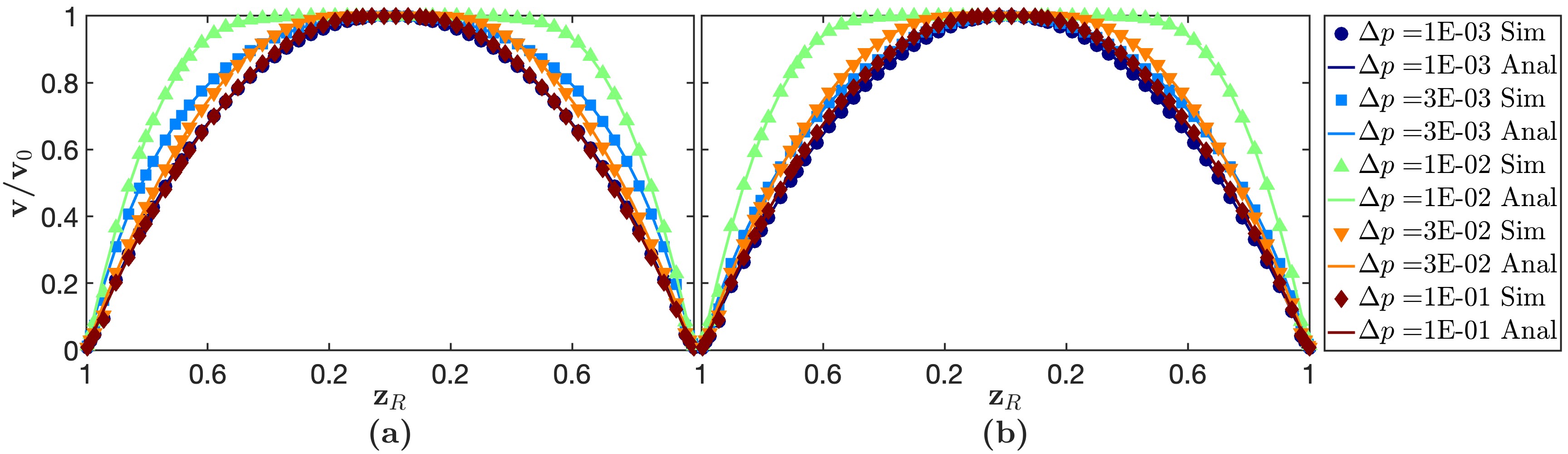}
    \caption{Velocity profiles for (a) Cross and (b) Carreau Fluid 4 flowing through a capillary tube. Pressure drops indicated in the legend have units of Pa.} 
    \label{fig:cap_vel_fluid4}
\end{figure}

\begin{figure}[t]
   \centering
    \includegraphics[width=1\linewidth]{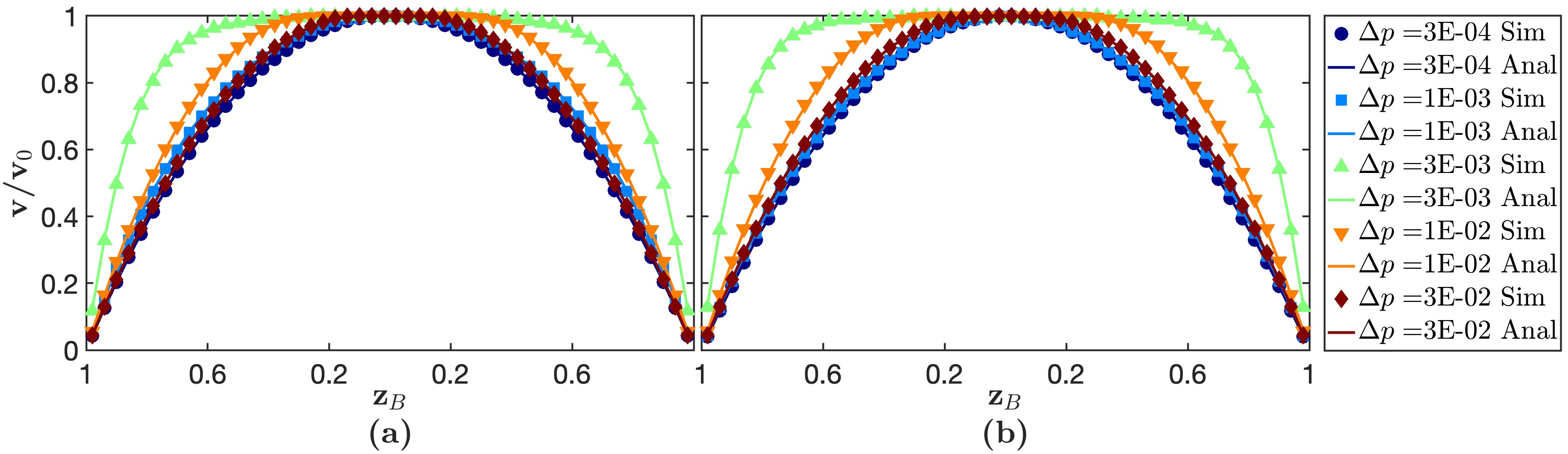}
    \caption{Velocity profiles for (a) Cross and (b) Carreau Fluid 4 flowing through a slit system. Pressure drops indicated in the legend have units of Pa.} 
    \label{fig:slit_vel_fluid4}
\end{figure}

The velocity profiles from microscale simulations were tabulated across a line that ran solely in the z-direction and passed through the centrum of each system, as described in detail in \S\ref{sec:methods}, and these were compared to velocities computed from Eqns (\ref{eq:vwr2}) through (\ref{eq:Iv_Carreau}). The microscale simulations (Sim) and analytical computations (Anal) are shown in Figures \ref{fig:cap_vel_fluid4} and \ref{fig:slit_vel_fluid4} for Fluid 4 flowing in the capillary tube and slit configurations respectively. In those figures the velocities were normalized by the maximum velocity for that pressure drop and the normalized values are plotted on the ordinate, while normalized distances from the centrum are plotted on the abscissa, $z_R$ for the capillary tube case and $z_B$ for the slit case. Results for Fluid 4 are displayed here because these results deviated the most from the Newtonian limit, which is the innermost velocity profile for each simulation. The results displayed were selected to show the transition from the Newtonian limit to the power-law regime and the greatest deviation from that limit. Simulation results for similarly selected runs are shown for Fluids 1--3 in supplementary material Figures S1--3 for the capillary tube case and Figures S4--6 for the slit case.  

\begin{figure}[t]
   \centering
    \includegraphics[width=\linewidth]{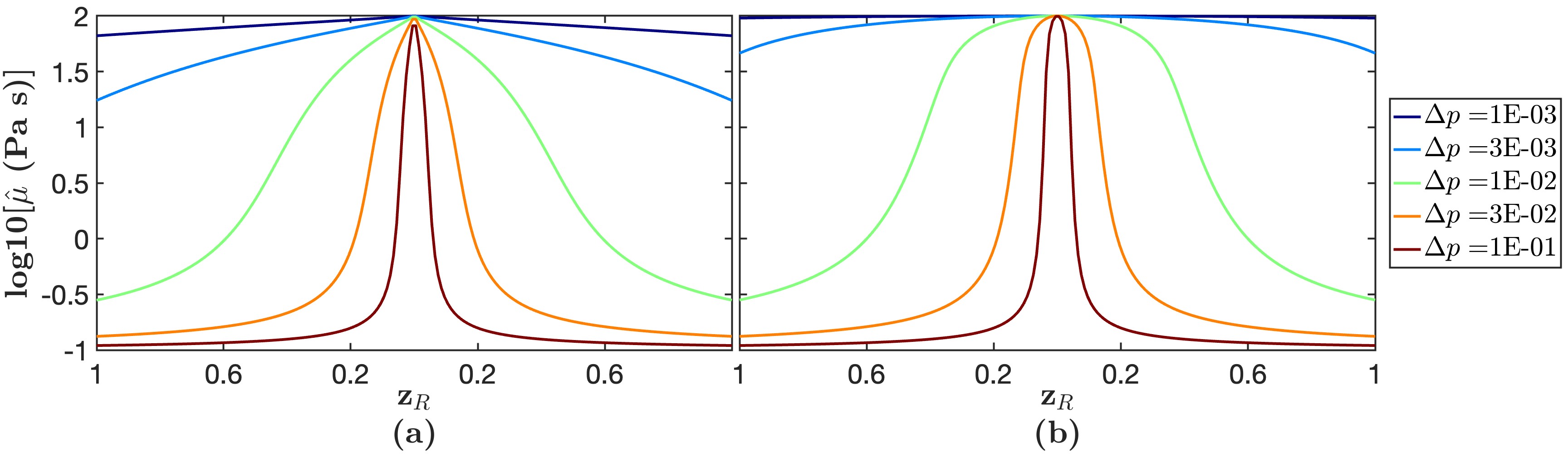}
    \caption{Viscosity profiles for (a) Cross and (b) Carreau Fluid 4 flowing through a capillary tube. Pressure drops indicated in the legend have units of Pa.} 
    \label{fig:cap_visc_fluid4}
\end{figure}

It may be observed that the velocity profile flattens out as the flow deviates from the Newtonian limit. After the maximum deviation from Newtonian flow is reached, the velocity profile gradually shifts back towards the Newtonian limit. This flattening of the velocity profile is due to the viscosity profile that develops throughout the cross-section of the systems, as shown in Figure \ref{fig:cap_visc_fluid4} for Fluid 4 in the capillary tube configuration. The maximum zero-shear viscosity is always achieved at the centrum of the system, decreasing to a minimum viscosity for that pressure drop at the system walls. At large pressure drops the viscosity profile tends toward the shape of a Dirac delta function, being at the infinite shear viscosity limit throughout the cross-section with the exception of the centrum where the viscosity is $\hsmu_0$.

\begin{figure}[t]
   \centering
    \includegraphics[width=1\linewidth]{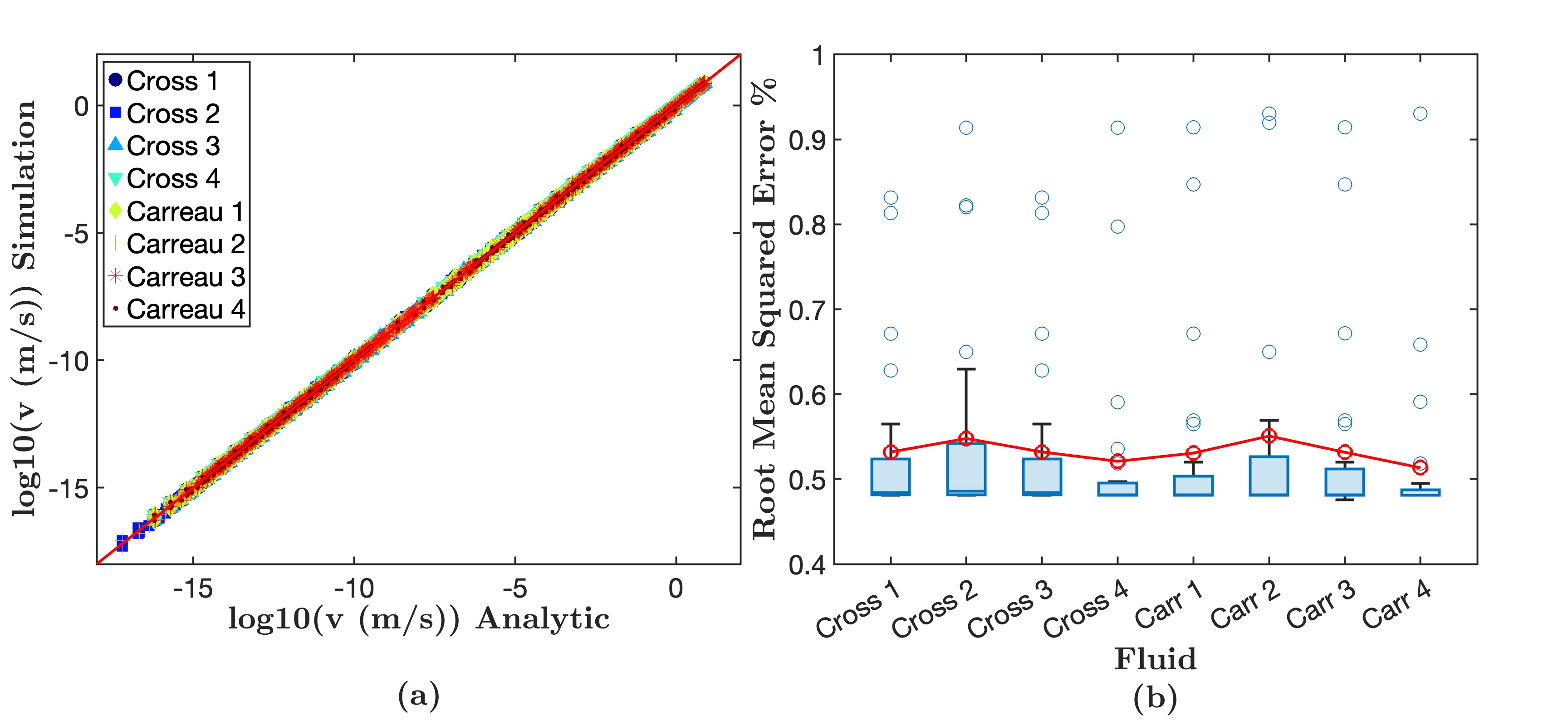}
    \caption{Velocity data for a capillary tube showing (a) predicted versus observed values and (b) box-and-whisker plots for the root-mean-squared-error percent. In the box-and-whisker plots red lines and markers show the algebraic average and blue markers show data that lies outside 1.5 times the inter-quartile range.} 
    \label{fig:cap_vel_error}
\end{figure}

\begin{figure}[t]
   \centering
    \includegraphics[width=1\linewidth]{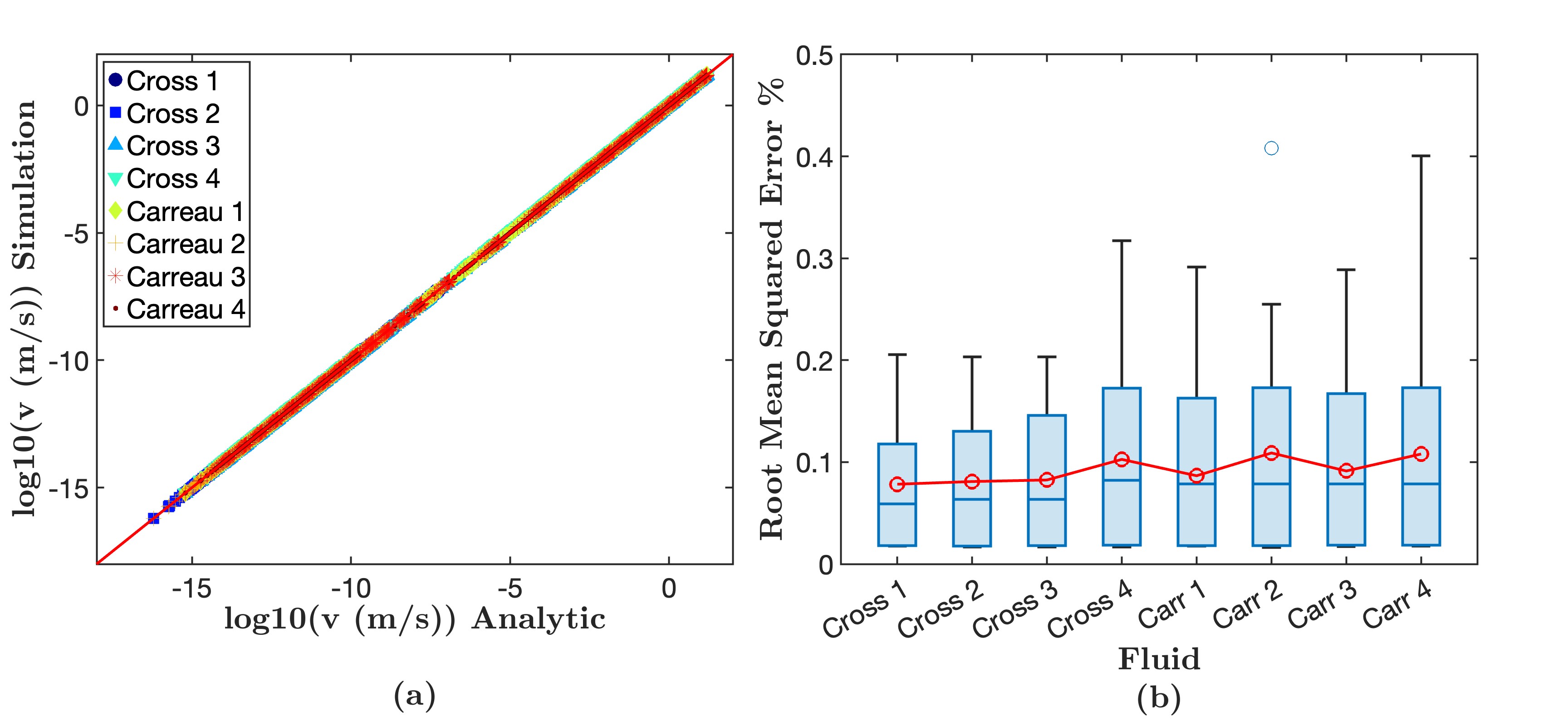}
    \caption{Velocity data for parallel plates showing (a) predicted versus observed values and (b) box-and-whisker plots for the root-mean-squared-error percent. In the box-and-whisker plots red lines and markers show the algebraic average and blue markers show data that lies outside 1.5 times the inter-quartile range.} 
    \label{fig:slit_vel_error}
\end{figure}

The RMSE box-plots and predicted versus simulated plots are presented in Figures \ref{fig:cap_vel_error} and \ref{fig:slit_vel_error}. These data show that there were many more points outside the IQR for the capillary tube than for the slit configuration, and that these existed for all fluids simulated in nearly equal measure. All of these errors occurred during the pressure drops where flow differed the most from the Newtonian limit, while this was not an issue for the slit configuration, indicating that they are due to the way in which the curved surface of the capillary tube is handled by OpenFOAM. Additional simulations were carried out with greater grid refinement, and it was observed that these errors decreased. Regardless of the outlying points, Figures \ref{fig:cap_vel_error} and \ref{fig:slit_vel_error} show close agreement between the analytical solution and microscale simulation, with even the greatest RMSEs being less than 1\%, and most simulations being within the expected error for these simulations based on the grid-refinement study, which was less than 0.5\%. Based on the low error present in the velocity simulations, the solutions derived during transport are validated next. 

\subsection{Advection Dominated Species Transport}

\begin{figure}[t]
   \centering
    \includegraphics[width=1\linewidth]{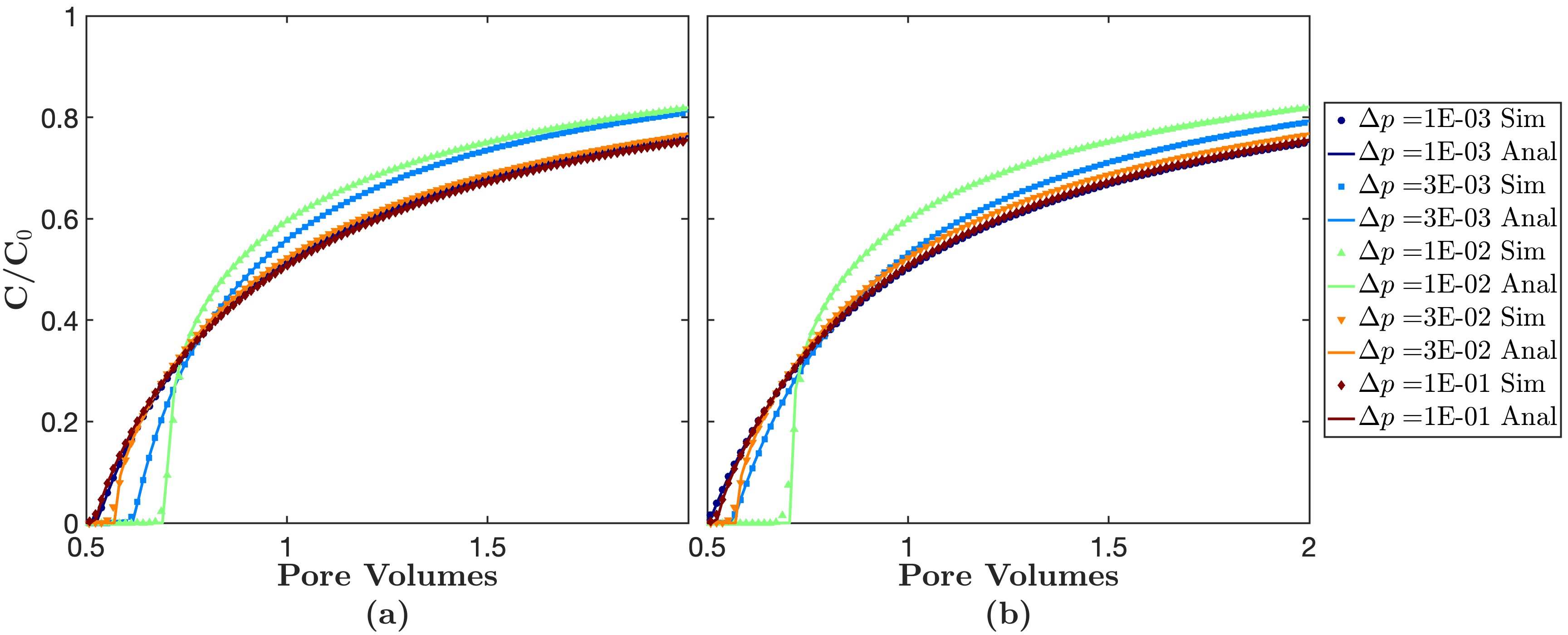}
    \caption{Average effluent concentration for (a) Cross and (b) Carreau Fluid 4 flowing through a capillary tube during advection dominated species transport, Pe = $10^5$. Pressure drops indicated in the legend have units of Pa. } 
    \label{fig:cap_HiPe_fluid4}
\end{figure}

Advection-dominated species transport occurs at high enough Pe such that diffusion is insignificant. Numerical transport simulations were conducted using the microscale velocity field output from flow simulations as described in \S \ref{sec:methods}. The average outlet concentration was computed from these numerical transport simulations and compared to the analytical solutions given by \Eqnstwo{CiA_analyticCap}{CiA_analyticSlit}, computed using the analytical velocity solutions. Plots for the normalized effluent concentration for the same pressure drops as were displayed in the velocity profile figures are presented in Figures \ref{fig:cap_HiPe_fluid4} and \ref{fig:slit_HiPe_fluid4} for Fluid 4 at Pe = $10^5$, with all other fluids presented in supplementary material Figures S7--12. These figures show the effluent concentration normalized by the influent concentration on the ordinate and the number of pore-volumes passed through the system plotted on the abscissa. 

\begin{figure}[t]
   \centering
    \includegraphics[width=1\linewidth]{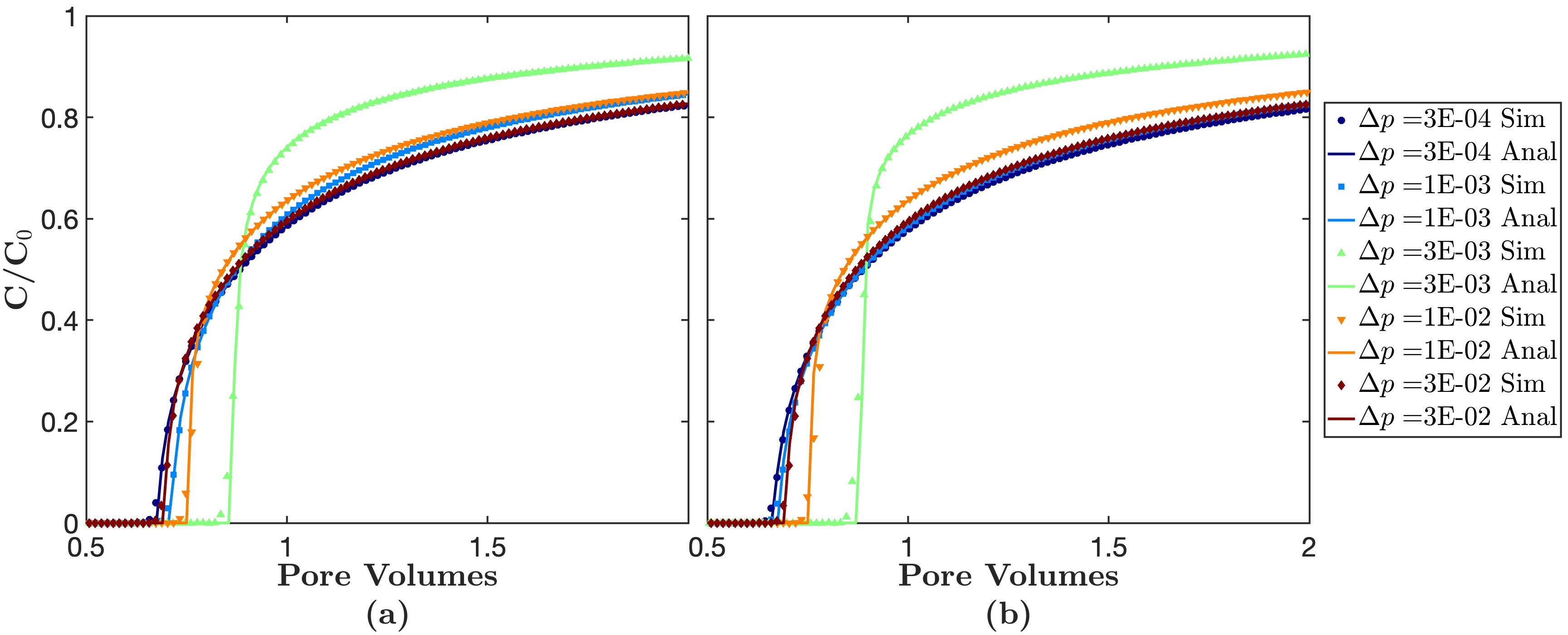}
    \caption{Average effluent concentration for (a) Cross and (b) Carreau Fluid 4 flowing through a parallel plate system during advection dominated species transport, Pe = $10^5$. Pressure drops indicated in the legend have units of Pa.} 
    \label{fig:slit_HiPe_fluid4}
\end{figure}

At low Pe the simulated effluent concentration differed significantly from the analytical solution, with RMSE > 10\%, until Pe = $10^5$. An additional set of simulations were carried out for Carreau Fluid 4 at Pe = $5\times 10^5$, and it was observed that these results differed from the Pe = $10^5$ results by less than 0.1\%, thus it follows that the advection dominated regime occurs at Pe $\geq$ $10^5$.

\begin{figure}[t]
   \centering
    \includegraphics[width=1\linewidth]{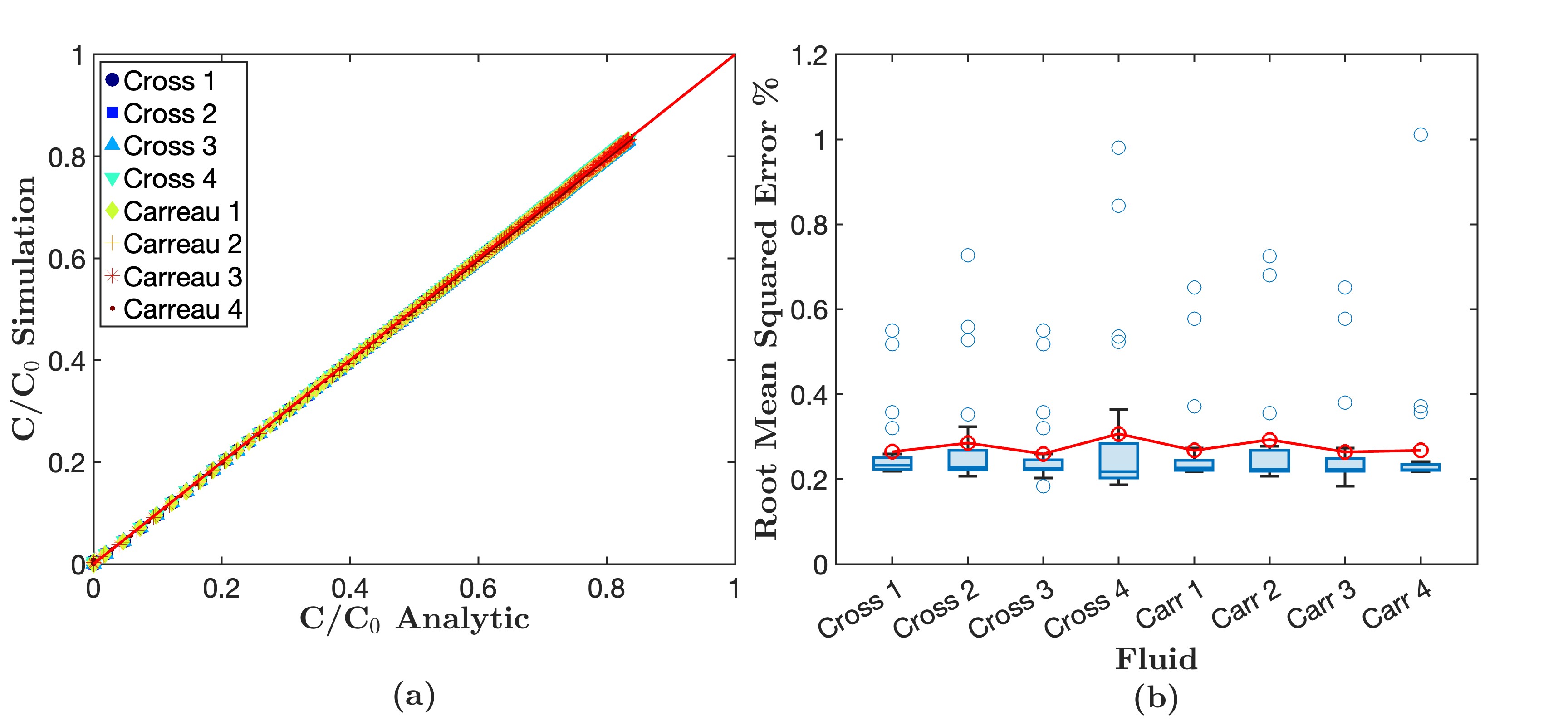}
    \caption{Data  for a capillary tube during high Pe transport (Pe = $10^5$) showing (a) predicted versus observed values and (b) box-and-whisker plots for the root-mean-squared-error percent. In the box-and-whisker plots red lines and markers show the algebraic average and blue markers show data that lies outside 1.5 times the inter-quartile range.} 
    \label{fig:cap_HiPe_error}
\end{figure}

\begin{figure}[t]
   \centering
    \includegraphics[width=1\linewidth]{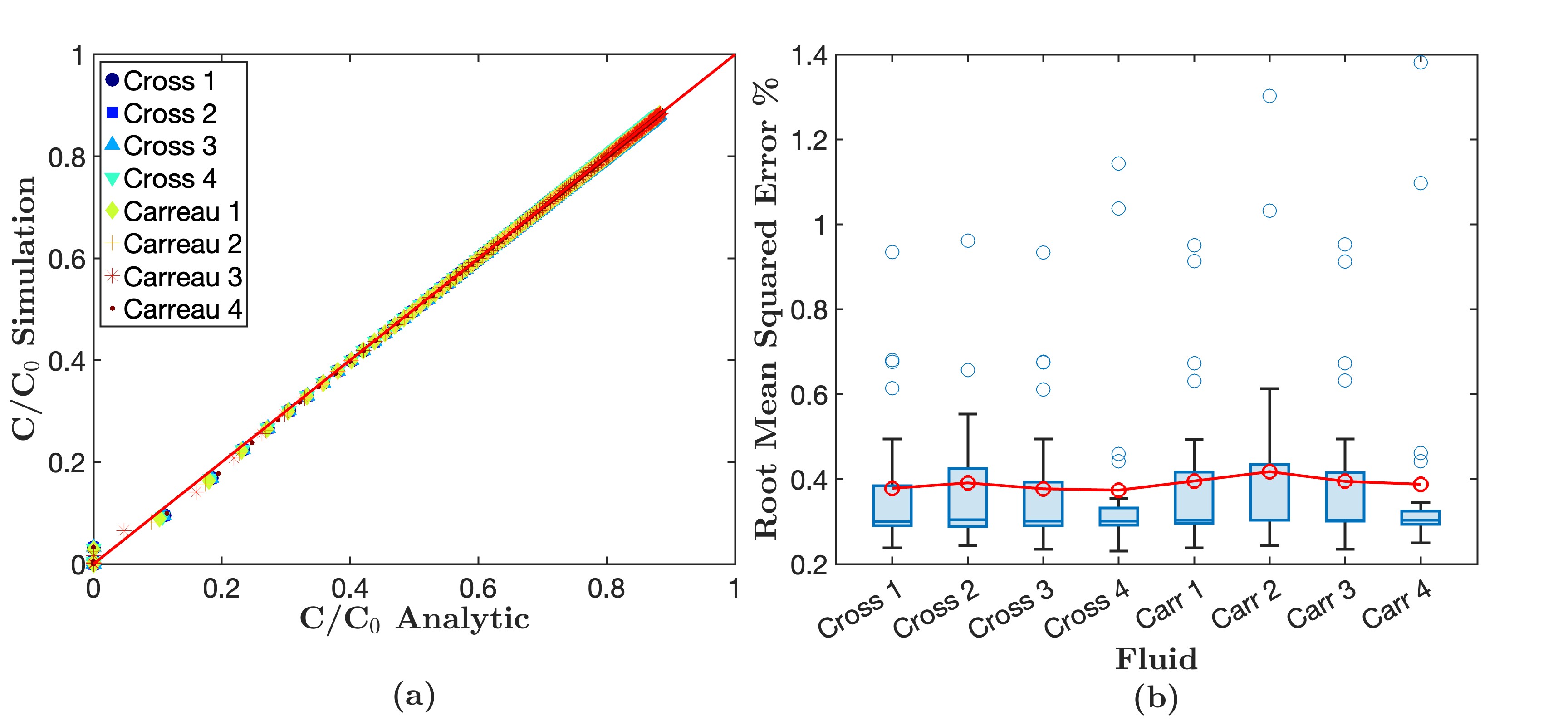}
    \caption{Data for the slit configuration during high Pe transport (Pe = $10^5$) showing (a) predicted versus observed values and (b) box-and-whisker plots for the root-mean-squared-error percent. In the box-and-whisker plots red lines and markers show the algebraic average and blue markers show data that lies outside 1.5 times the inter-quartile range.} 
    \label{fig:slit_HiPe_error}
\end{figure}

The analytical solutions generally exhibited error of less than 1\% based on averaged microscale simulation, showing agreement within the numerical error of the simulations. Similarly to the velocity solution validation, predicted-versus-actual and box-and-whisker plots were generated from the data to show the agreement for each system, presented in Figures \ref{fig:cap_HiPe_error} and \ref{fig:slit_HiPe_error} for capillary tube and slit systems respectively. The box-plots were generated using the 500 transport simulations carried out for Pe = $10^5$.

\begin{figure}[t]
   \centering
    \includegraphics[width=1\linewidth]{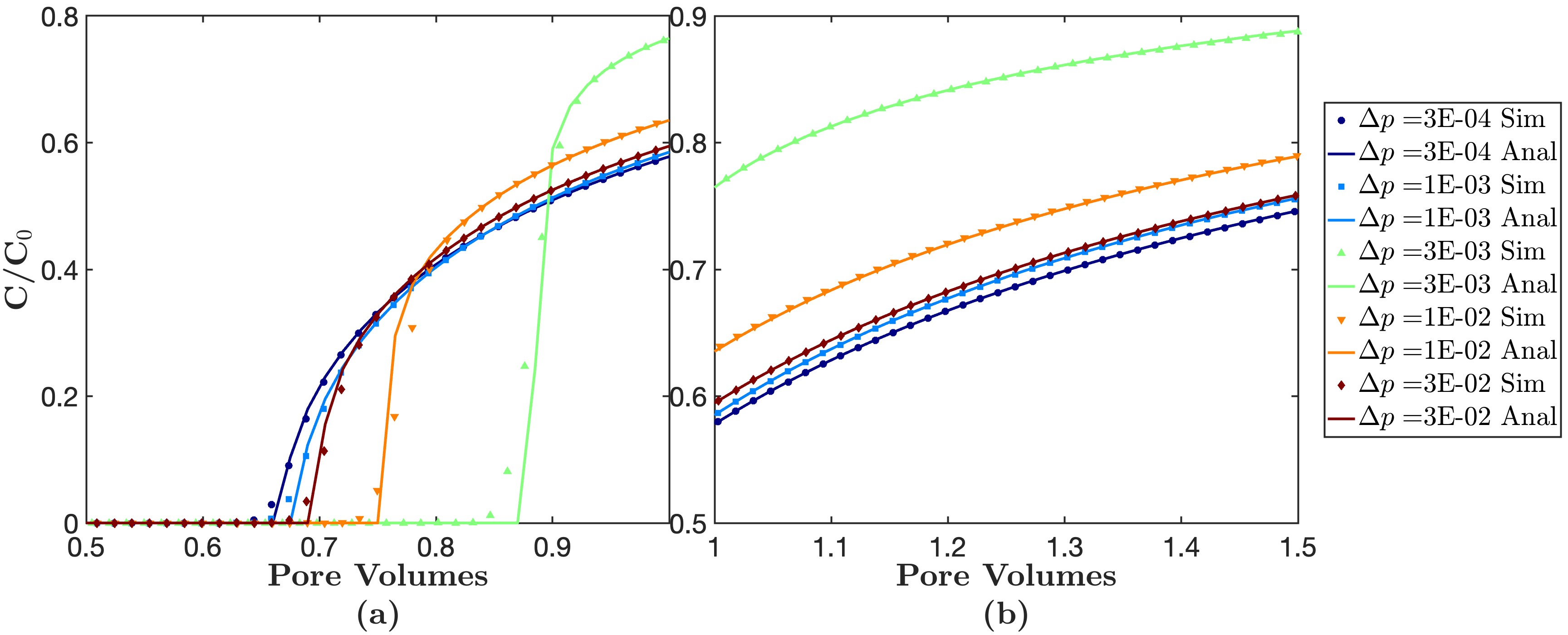}
    \caption{Average effluent concentration for Carreau model Fluid 4 flowing through a parallel plate system during advection dominated species transport (Pe = $10^5$) scaled from (a) 0.5--1 and (b) 1--1.5 pore-volumes. Pressure drops indicated in the legend have units of Pa.} 
    \label{fig:slit_HiPe_fluid4_zoom}
\end{figure}

While the IQR for each system lay below the 0.5\% expected for these simulations, several outliers can be seen that reach 1\%, especially in the case of the slit configuration. This is notable as the slit exhibited less deviation between the microscale simulations and the analytical velocity solution, as can be seen in Figure \ref{fig:slit_vel_error}. One explanation for this becomes apparent when comparing the breakthrough curves in Figures \ref{fig:cap_HiPe_fluid4} and \ref{fig:slit_HiPe_fluid4}. It can be seen that the breakthrough curves are sharper for the slit configuration than they are for the capillary tube. Sharp fronts present a known challenge during numerical simulation \cite{Yang_Samper_09,Wang_Dai_etal_12}, which can lead to numerical approximation errors. A rescaled image of Carreau model Fluid 4 is presented in Figure \ref{fig:slit_HiPe_fluid4_zoom} at Pe = $10^5$, where the microscale simulation data can be seen to deviate from the analytical solution most as the sharp front interacts with the outlet of the system before matching the analytical solution closely after one pore-volume has passed through. This is also clear in Figure \ref{fig:slit_HiPe_error}(a) where the predicted concentrations deviated most from the observed concentrations at values below $C/C_0=0.2$. A simulation was carried out after doubling the mesh refinement in each dimension and the error was observed to decrease, indicating that the sharp front errors could be reduced with high mesh refinement. 

\subsection{Enhanced Molecular Diffusion}

\begin{figure}[t]
   \centering
    \includegraphics[width=1\linewidth]{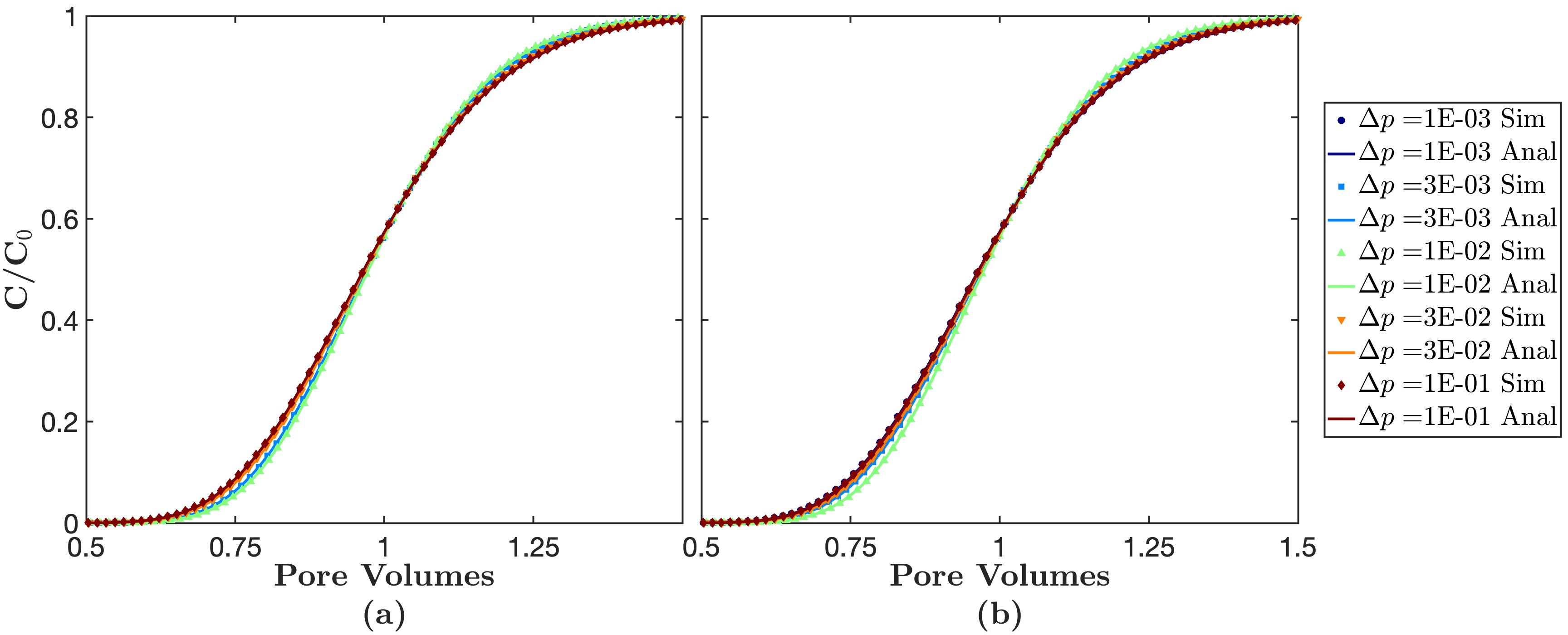}
    \caption{Average effluent concentration for (a) Cross and (b) Carreau Fluid 4 flowing through a capillary tube during enhanced molecular diffusion, Pe = $10^2$. Pressure drops indicated in the legend have units of Pa.} 
    \label{fig:cap_tp_fluid4}
\end{figure}

\begin{figure}[t]
   \centering
    \includegraphics[width=1\linewidth]{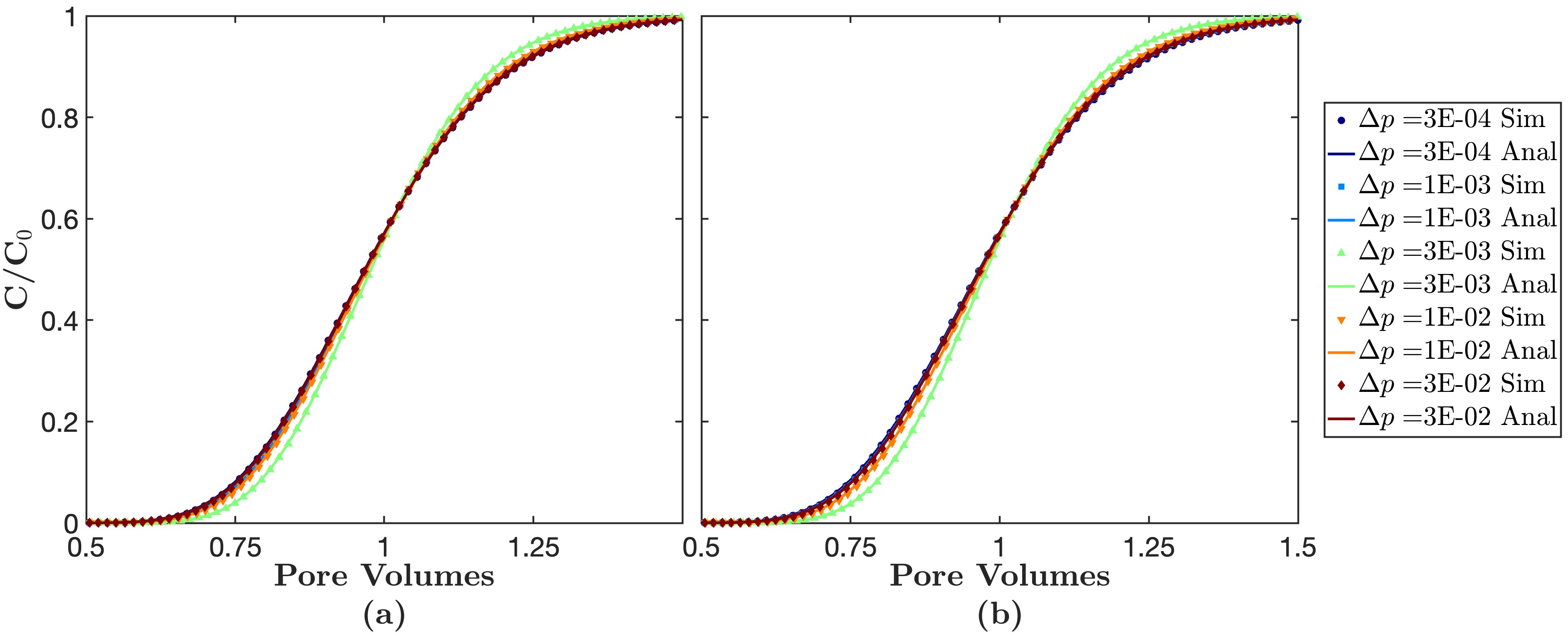}
    \caption{Average effluent concentration for (a) Cross and (b) Carreau Fluid 4 flowing through a parallel plate system during enhanced molecular diffusion, Pe = $10^2$. Pressure drops indicated in the legend have units of Pa.} 
    \label{fig:slit_tp_fluid4}
\end{figure}

The enhanced molecular diffusion regime occurs at low enough Pe that diffusion is the dominant mechanism for transport in directions perpendicular to flow while advection dominates in the flow direction, leading to a species front that deforms based on the velocity profile. The average outlet concentration computed from microscale simulations were compared to macroscale transport modeling data that resulted from using the enhanced molecular diffusion coefficients derived from \Eqnstwo{hsDwCapF}{hsDwSlitF}. The microscale simulation and macroscale model data are presented for Fluids 4 at Pe = $10^2$ in Figures \ref{fig:cap_tp_fluid4} and \ref{fig:slit_tp_fluid4} for the capillary tube and slit configurations respectively, with normalized concentration and pore-volumes plotted similarly to the high Pe case. Figures showing the same data plotted for Fluids 1--3 are presented in supplementary material Figures S13--18. 

The RMSE between averaged microscale simulations and macroscale modeling exceeded 10\% until Pe $\leq$ 100, with the onset of non-Fickian transport characteristics becoming evident at higher Pe. This provides an upper bound for the enhanced molecular diffusion regime, while there is no theoretical lower Pe bound for the regime as the enhanced molecular diffusion coefficients derived from \Eqnstwo{hsDwCapF}{hsDwSlitF} approach the molecular diffusion coefficient as Pe approaches zero. 

\begin{figure}[t]
   \centering
    \includegraphics[width=1\linewidth]{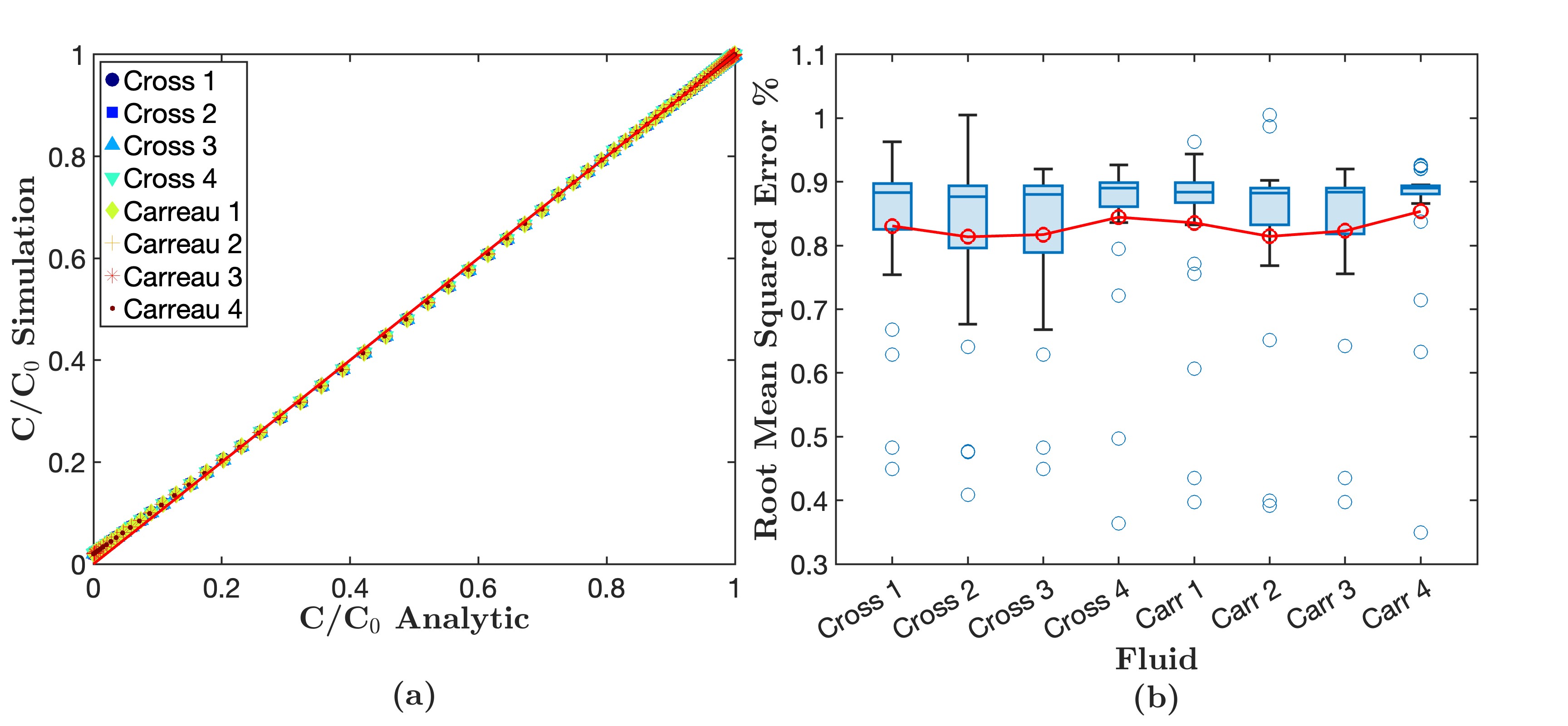}
    \caption{Data  for a capillary tube during enhanced molecular diffusion regime transport (Pe = $10^2$) showing (a) predicted versus observed values and (b) box-and-whisker plots for the root-mean-squared-error percent. In the box-and-whisker plots red lines and markers show the algebraic average and blue markers show data that lies outside 1.5 times the inter-quartile range.} 
    \label{fig:cap_tp_error}
\end{figure}

\begin{figure}[t]
   \centering
    \includegraphics[width=1\linewidth]{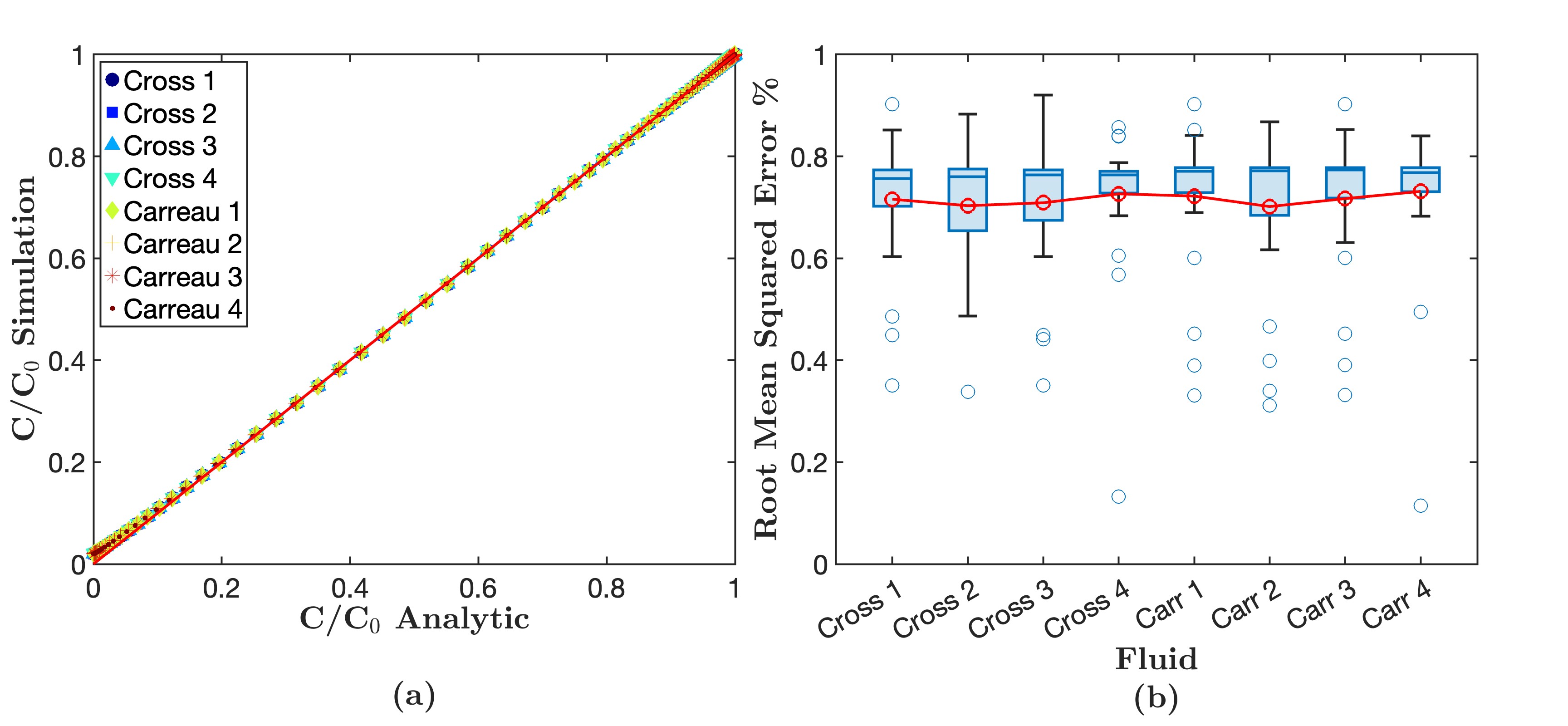}
    \caption{Data for the slit during enhanced molecular diffusion regime transport (Pe = $10^2$) showing (a) predicted versus observed values and (b) box-and-whisker plots for the root-mean-squared-error percent. In the box-and-whisker plots red lines and markers show the algebraic average and blue markers show data that lies outside 1.5 times the inter-quartile range.} 
    \label{fig:slit_tp_error}
\end{figure}

As for the velocity and high Pe validations, prediction-versus-observation and box-and-whisker plots were generated for the enhanced molecular diffusion transport simulations, presented in Figures \ref{fig:cap_tp_error} and \ref{fig:slit_tp_error} for the capillary tube and slit configurations, respectively. These data were generated using the results for Pe = 100, resulting in 100,000 points of comparison. All simulations exhibited less than 1\% RMSE, but exhibited more error than the velocity and high Pe transport simulations. This may be expected, as several assumptions were made to derive the solutions for the enhanced molecular diffusion coefficients. Investigation into each of these assumptions and their impact on the errors present may be a focus of future work. 

\begin{figure}[t]
   \centering
    \includegraphics[width=1\linewidth]{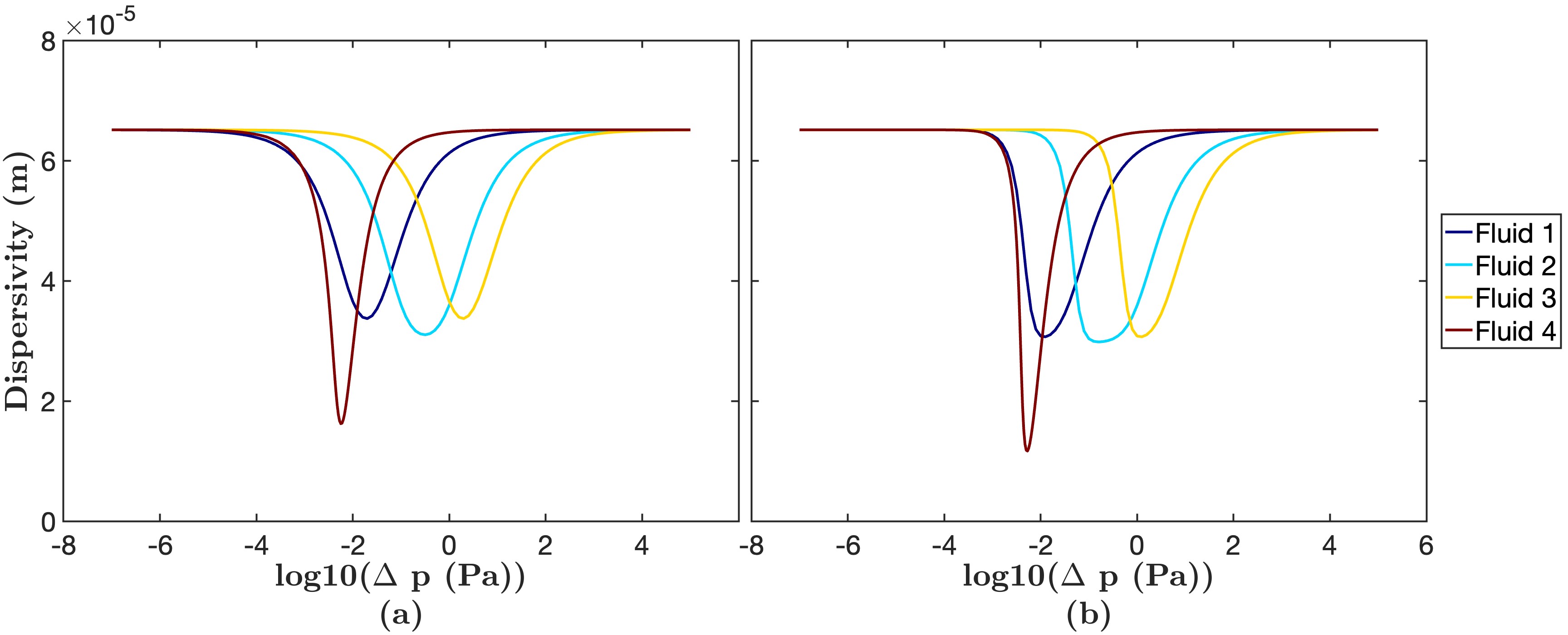}
    \caption{Dispersivity for (a) Cross and (b) Carreau Fluid 4 flowing through a capillary tube during enhanced molecular diffusion, Pe = $10^2$.} 
    \label{fig:cap_disp}
\end{figure}

\begin{figure}[t]
   \centering
    \includegraphics[width=1\linewidth]{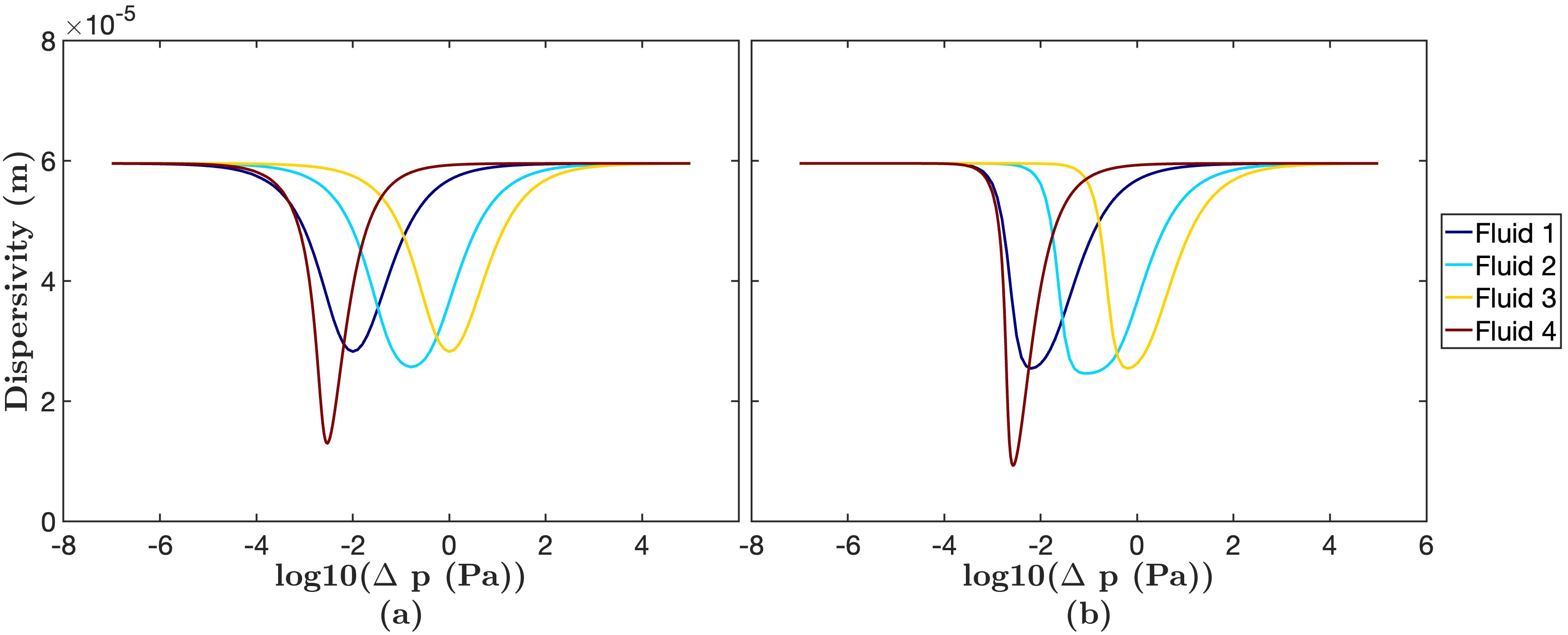}
    \caption{Dispersivity for (a) Cross and (b) Carreau Fluid 4 flowing through a parallel plate system during enhanced molecular diffusion, Pe = $10^2$.} 
    \label{fig:slit_disp}
\end{figure}

The breakthrough curves presented in Figures \ref{fig:cap_tp_fluid4} and \ref{fig:slit_tp_fluid4} exhibited a shift to a sharper front similar to the high Pe case. To make this clear, the dispersivities computed from \Eqn{hasawL} for each simulation are plotted in Figures \ref{fig:cap_disp} and \ref{fig:slit_disp} versus log10 of the pressure drop for the system. The dispersivity values decreased by a factor of six in the most extreme cases, and a factor of two for most fluids. This is in contrast to what has been observed for sphere packings where dispersivity increased as GNF flow deviated from the Newtonian limits\cite{Bowers_Miller_25}. This is also a more pronounced shift from Newtonian dispersivity than has been shown in past work. 

An explanation for the difference between dispersivity deviation for the systems investigated here and the random sphere packings in Bowers and Miller (2025)\cite{Bowers_Miller_25} is that only the dispersivity limiting behavior of GNF flow is present here, while in random sphere packings there are two competing mechanisms. These two mechanisms are the flattening of velocity profiles across pore-throats, as depicted here in Figures \ref{fig:cap_vel_fluid4} and \ref{fig:slit_vel_fluid4}, as well as the increase in tortuosity that has become evident during GNF flow in unstructured porous media\cite{Zhang_Prodanovic_etal_19,Bowers_Miller_25}.

\section{Discussion}

Capillary tubes and parallel plates present idealized systems that many have found to be useful when investigating porous media \cite{Taylor_53,Sirs_91,Rana_Liao_19,Kutev_Tobakova_etal_21,Boschan_Ippolito_etal_08,Sochi_15,Zhang_Prodanovic_etal_19}. This has been especially the case for non-Newtonian fluid flow, which presents fluid mechanical complexity enough from their rheology without introducing the additional intricacies of real porous media \cite{Malvault_Ahmadi_etal_17,Castro_19,Castro_Goyeau_21}. While the analytical transport solutions derived here were applied to GNFs, they have been derived generally and can be applied to other non-Newtonian flows as long as velocity data is available. Such velocity data can be generated from an analytical solution such as was done here, or from numerical simulation, negating the need to carry out computationally expensive transport simulations.

The general transport solutions derived here could be used to develop first pass simulators for viscoplastic and viscoelastic fluids that have heretofore been difficult to model \cite{Roustaei_Chevalier_etal_16,Bauer_Talon_etal_19,Chaparian_Izbassarov_etal_20,Kumar_Aramideh_etal_21}. In particular there are few numerical methods available to simulate viscoelastic fluids \cite{Favero_Secchi_etal_10,De_Das_etal_16,Browne_Shih_etal_20}, although velocity profiles could be generated from imaging \cite{Browne_Shih_etal_20,Browne_Datta_21,Browne_Datta_24} that could be used to carry out the transport computations done using a semi-analytical solution here. The analytical transport solutions could then be used to develop pore-network models which, while limited, are computationally efficient and common in the literature \cite{Sorbie_Clifford_etal_89,Perrin_Tardy_etal_06,Dong_Blunt_09,Sochi_10,Fayed_Sheikh_etal_16,Basset_Najm_etal_19,Hauswirth_Najm_etal_19,Castro_Goyeau_21,Suo_Foroughi_etal_25}. Based on these pore-network models, physical understanding can begin to be developed as more advanced numerical methods are introduced. 

An interesting observation of the data presented in this work is that dispersive transport phenomena decreased as GNF flow behavior increased. This can be observed in the high Pe simulations where the species front sharpened, as well as in the change in dispersivity observed during enhanced molecular diffusion. The flattening out of the velocity profiles in both the capillary tube and slit systems explain this phenomena, leading to a narrower distribution in velocity values and thus less dispersion. This contrasts with an increase in dispersion that has been observed for transport during GNF flow in more complex porous media, such as random sphere packings \cite{Bowers_Miller_25}. The disparity indicates that there are potentially two competing mechanisms that impact transport during non-Newtonian flow, one related to the velocity profile observed across individual pores, and one related to the distribution of velocities throughout a complex porous medium. Future work will need to be done to determine how these two competing mechanisms interact with one another, and when one can be expected to dominate over the other.

\section{Summary and Conclusions}

Semi-analytical solutions for the fluid velocity in capillary tubes and slits were derived for Cross and Carreau model fluids, which are common classes of generalized Newtonian fluids. It was then shown how these velocity solutions could be used to compute the average concentration across the cross section of both geometries during advection dominated flow. An analytical solution for cross-sectionally averaged macroscale concentration and the effective diffusion coefficient during the enhanced molecular diffusion flow regime were then derived for both geometries. These general analytical solutions were then used to simulate species transport in both capillary tubes and slits during Cross and Carreau model fluid flow using the semi-analytical solutions for velocity derived for the corresponding geometries.

The flow and transport solutions were compared to microscale simulations carried out using OpenFOAM, an open-source simulation toolkit. The velocity solutions were within the numerical error of the microscale simulations, and exhibited a flattened out profile due to the viscosity profiles that developed for each fluid class. Semi-analytical species transport solutions were also within the numerical error of the microscale simulations, and exhibited less dispersion in both the capillary tube and slit geometries compared to Newtonian flow. To quantify the change in dispersive phenomena for the generalized Newtonian fluids compared to a Newtonian fluid, the dispersivity was computed for the enhanced molecular diffusion case and it was shown that dispersivity dropped by up to a factor of six depending on the flow rate and rheological properties of the fluids. 

The analytical solutions developed here for species transport, while applied to Cross and Carreau model fluids, were not derived for any specific fluid rheology. Thus the solutions derived herein can be applied for transport in other generalized Newtonian fluids and even for cases where only experimental velocimetry observations are available and the underlying rheology of the fluid has not been characterized.  

\section*{Supplementary Material}
Supplementary material has been provided which includes: a table of pressure drops used during simulations (Table SI); velocity profiles for Fluids 1-3 in the capillary tube and slit geometries (Figures S1-6); average effluent concentration breakthrough curves during high Pe flow for Fluids 1-3 in the capillary tube and slit geometries (Figures S7-12); and average effluent concentration breakthrough curves during enhanced molecular diffusion for Fluids 1-3 in the capillary tube and slit geometries (Figures S13-18).

\begin{acknowledgments}
This work was supported by National Institute of General Medical Sciences grant R41GM136084. Christopher A. Bowers would like to acknowledge XCMR Inc. for providing financial support.
\end{acknowledgments}

\section*{Author Declarations}

\subsection*{Conflict of Interest}
\noindent The authors have no conflicts to disclose.

\subsection*{Author Contributions}
\noindent \textbf{Christopher Bowers:} Conceptualization (equal); Methodology (lead); Software (lead); Visualization (lead); Writing -- original draft (lead); Writing -- review \& editing (equal). \textbf{Cass Miller:} Conceptualization (equal); Resources (lead); Supervision (lead); Writing -- original draft (supporting); Writing -- review \& editing (equal).

\section*{Data Availability}
\noindent The data that support the findings of this study are available from the corresponding author upon reasonable request.

\bibliography{PoF_bib,gwatername}

\end{document}